\documentstyle[11pt]{article}

\newcommand{\sect}[1]{\setcounter{equation}{0}\section{#1}}
\newcommand{\subsect}[1]{\subsection{#1}}

\newcommand{\vs}[1]{\rule[- #1 mm]{0mm}{#1 mm}}

\newcommand{\lbl}[1]{\label{eq:#1}}
\newcommand{\rf}[1]{(\ref{eq:#1})}
\newcommand{\nn}{\nonumber} 
\newcommand{\be}{\vs{2}\begin{equation}}
\newcommand{\ee}{\vs{2}\end{equation}}

\newcommand{\bea}{\begin{eqnarray}}
\newcommand{\ena}{\end{eqnarray}}
\newcommand{\nnbea}{\begin{eqnarray*}}
\newcommand{\nnena}{\end{eqnarray*}}
\newcommand{\leqa}{\lefteqn}

\newcommand{\lra}{\ \longrightarrow\ }

\newcommand{\ovl}[1]{\overline{#1}}
\renewcommand{\bar}{\ovl}

\newcommand{\sm}[2]{\textstyle{\frac{#1}{#2}}\displaystyle}

\newcommand{\dps}{\displaystyle}

\newcommand{\bz}{{\bar{z}}}

\newcommand{\bZ}{{\bar{Z}}}
\newcommand{\yz}{y_{z}}
\newcommand{\ybz}{\bar{y}_{\bz}}

\newcommand{\YZ}{Y_{Z}}

\newcommand{\YbZ}{\bar{Y}_{\bZ}}

\newcommand{\zbz}{(z,\bz)}
\newcommand{\zy}{{(z,y)}}

\newcommand{\zY}{{(z,Y)}}
\newcommand{\yZ}{{(z,Y)}}
\newcommand{\ZY}{{(Z,Y)}}

\newcommand{\YbY}{(\YZ,\YbZ)}

\newcommand{\cP}{{\cal P }} 

\newcommand{\cU}{{\cal U }}
\newcommand{\sS}{{\cal S }}
\newcommand{\cK}{{\cal K }}
\newcommand{\cC}{{\cal C }}

\newcommand{\cA}{{\cal A }}
\newcommand{\cD}{{\cal D }}
\newcommand{\cL}{{\cal L }}
\newcommand{\cbD}{\bar{\cal D }}
\newcommand{\cT}{{\cal T }}
\newcommand{\cM}{{\cal M }}
\newcommand{\cS}{{\cal S }}
\newcommand{\cB}{{\cal B }}

\newcommand{\cZ}{{\cal Z }}

\newcommand{\cQ}{{\cal Q }}
\newcommand{\cF}{{\cal F}} 

\newcommand{\cJ}{{\cal J }}

\newcommand{\prt}{\partial}
\newcommand{\prtz}{\partial_z}
\newcommand{\prtbz}{\bar{\partial}_{\bar{z}}}

\newcommand{\lambdan}{\lambda_z^{\Zn}}
\newcommand{\blambdan}{\bar{\lambda}_\bz^{\Zn}}

\newcommand{\lambdapi}{\lambda_z^{Z^{(p_i)}}}
\newcommand{\prtZ}{{\partial}_{ Z}}

\newcommand{\prtY}{\frac{\partial}{\partial Y_Z}}
\newcommand{\prtbY}{\frac{\partial}{\partial \bar{Y}_{\bZ}}}
\newcommand{\bprt}{\bar{\partial}}
\newcommand{\blambda}{\bar{\lambda}}
\newcommand{\mub}{\bar{\mu}}
\newcommand{\Zn}{Z^{(n)}}
\newcommand{\Zr}{Z^{(r)}}
\newcommand{\Zbn}{\bZ^{(n)}}
\newcommand{\Zbr}{\bZ^{(r)}} 

\newcommand{\ZBZ}{(Z,\bar{Z})}
\newcommand{\ZBZn}{(Z^{(n)},\bar{Z}^{(n)})}
\newcommand{\ZBZr}{(Z^{(r)},\bar{Z}^{(r)})}

\newcommand{\zbzp}{(z',\bar{z'})}
\newcommand{\zbzpp}{(z",\bar{z"})}

\newcommand{\z}{(z)}

\newtheorem{guess}{Theorem}[section]
\newtheorem{statement}{Statement}[section] 
\begin{document}
\renewcommand{\thefootnote}{\fnsymbol{footnote}}
\pagestyle{empty}

\title{\bf The role of complex structures in $w$-symmetry}

\vskip 1.5cm

\author{G. Bandelloni\\
{\em Dipartimento di Fisica dell' Universit\`a di Genova}\\
{\em Istituto Nazionale di Fisica Nucleare, INFN, Sezione di Genova}\\
{\em via Dodecaneso 33, I-16146 GENOVA, Italy}\\
e-mail beppe@genova.infn.it\\[2mm] 
and\\[2mm]
S. Lazzarini\\
{\em Centre de Physique Th\'eorique}\\
{\em CNRS-Luminy, Case 907, F-132288 MARSEILLE Cedex 9, France}\\
e-mail sel@cpt.univ-mrs.fr }

\maketitle

\begin{abstract}
In a symplectic framework, the infinitesimal action of
symplectomorphisms together with suitable reparametrizations of the two
dimensional complex base space generate some type of
$W$-algebras. It turns out that complex structures parametrized by
Beltrami differentials play an important role in this
context. The construction parallels very closely
two dimensional Lagrangian conformal models where Beltrami
differentials are fundamental.
\end{abstract}

\vskip 1.5cm


\indent

\noindent {CPT-99/P.3924}, Internet : {\tt www.cpt.univ-mrs.fr}

\newpage
\pagestyle{plain}
\setcounter{page}{1}

\sect{Introduction}

In the last decade, $W$-algebras have  provided a unifying
landscape for various topics like integrable systems,
conformal field theory, as well as uniformization of 2-dimensional
gravity.
They were originally discovered as a natural extension of the 
Virasoro algebra by Zamolodchikov \cite{Zam} and later implicitly by
Drinfield and Sokolov \cite{DrinSok}. The latter obtained the classical
$W$-algebras by equipping the coefficients of first order matrix 
differential operators 
with the second Gel'fand-Dickey Poisson bracket \cite{GelDik}.

In the study of two dimensional conformal models in the so-called
conformal gauge $W$-gravity can be defined as a generalization of the
reparametrization invariance such that in the conformal gauge  
two copies of the corresponding $w$-algebra are obtained.
Moreover, it has be found that the matrix differential operator 
has to be supplemented by another equation which is usually
 referred as the {\it Beltrami equation} \cite{DrinSok}.
  
Several attempts have been made in order to give a geometric picture of this
very rich structure provided by $w$-algebras .
Various aspects of this issue have been tackled independently by many
authors and it is by now universally referred to as $w$-geometry
\cite{4,7,7a}. This geometry is naturally related to $W$-gravity. 

On the one hand, the geometric structure turns out to be related
to the uniformization of Riemann surfaces; for instance in
\cite{govin} a  uniformization in higher  dimensions was shown to
be related to the Teichm\"uller spaces  constructed by Hitchin
\cite{Hit}.  However, the Beltrami differentials which naturally
occur there are related to KdV flows but not to the complex
structures underlying the $w$-geometry, in contrast to what it
will be shown it the present paper.
   
On the other hand, Witten \cite{Wi1} and Hull \cite{Hull}
both pointed out that the use of Poisson bracket induces symplectic 
forms on certain manifolds and, in doing so, they proposed to study the
role of symplectic diffeomorphisms in the construction of the
$w$-algebras.
Symplectic diffeomorphisms (or symplectomorphisms) form a class of
diffeomorphisms on the cotangent bundle over the configuration space 
(phase space) which leave the canonical symplectic form invariant.  

This type of invariance is very rich in the sense that it is the
infinitesimal action of symplectomorphisms on a numerable set
(may be even infinite) of very peculiar smooth changes of
coordinates on the base Riemann surface which generates
$w$-algebras as it has be shown quite recently \cite{BaLa0}.
Similarly to the fact that the moduli space of Riemann surfaces
plays an important role in 2d-conformal field theory coupled to
gravity, it is expected that the same ought to hold for the
$w$-symmetry \cite{deBoerGoeree}.

 Our treatment gives in an explicit way the infinitesimal
  mappings, so we can automatically  give  a set of Beltrami
  parametrizations \cite{becchi} of them  which can represent this
  moduli space.    These problems, mostly topological, have an
  important counterpart within Lagrangian  Quantum Field Theory,
  involving locality: indeed the occurrence of a numerable set
  of{\it{ "Beltrami equations"}} imposes a non local dependence of
  some fields on other ones, spoiling the fundamental requirements
  of local Quantum Field Theory.  We do not realize the link
  between our reparametrizations  and the gauge transformations of
  the flat $SL(n,{\bf C})$  vector bundles canonically associated
  to the generalized  projective structures, as proposed by
  Zucchini \cite{Zucchini}, and the embedding transformations
  proposed in References \cite{24,25,26}, but we hope
  that the absolute general statement of our transformations
  could shed some new light on the geometrical nature of $w$-symmetry.

Other examples of $w$ algebras, where the relevance of the complex 
structure occurs, were considered in many References
\cite{Laz1,Laz2,Gieres,Sorella,Grimm,Ader1,Ader2}. The former
can be reconstructed by a partial breaking of the reparametrization
invariance, while a set of consistency conditions
controls the spoiling of the breaking of the symmetry under
reparametrizations \cite{BaLa0}.
                     
Moreover our approach differs slightly from these ones since it is
 grounded  on a set of complex structures  parametrized by
 Beltrami differentials, whose expressions give a geometrical
 meaning to quantities introduced in these References.

   We stress that our  philosophy relies on a deep connection
 between reparametrization invariance and $w$-symmetry, so if we
 want to investigate the physical  aspects of this problem, a
 correct geometrical formulation will prove to be of great help.

  In this context we consider a Quantum Extension of the theory,
  within the Lagrangian framework. However, we shall see that the locality
  (in the fields) assumption  
  of a common Lagrangian Theory cannot be fulfilled; anyhow 
  the dismission of this requirement will provide the 
  mechanism for the compensation of the Quantum Anomalies and 
  the improvement at the quantum level of physical quantities 
  (Energy Momentum tensors) with definite holomorphic properties 
involving well defined complex structures.   
  
Section 2 recalls some of the main results obtained in a previous work
\cite{BaLa0} with emphasis on the notation and the basic symplectic
geometry underlying our problem.

   In Section 3 we study the  geometrical properties of the
   spaces, whose mappings toward a common  background space define
   the transformations leading to $w$ symmetry.

 In Section 4 we find, within a B.R.S. framework, the $w$ algebra
 transformations.
  
 In Section 5 we build a very general Lagrangian Quantum Field
 theory, and using the deep connections between reparametrization
 invariance and $w$ symmetry  we improve its B.R.S. Quantum
 extension. An Appendix is devoted to the cohomological problems
 and  and Spectral Sequences calculations.

\bigskip
  
\sect{Geometrical Approach}

The starting point of a symplectic structure is the definition on a
manifold 
of the canonical 1-form on $T^{*}\zy$ $\theta$ defined in a  
local chart frame ${\cU_\zy}$

\begin{eqnarray}
{\theta\arrowvert}_{\cU_\zy}=\biggl[\yz dz  +\ybz d\bz  \biggr]
\lbl{thetaz}
\end{eqnarray}

The $(2,0)-(0,2)$ form 
$\Omega\equiv d\theta$

\begin{eqnarray}
{\Omega\arrowvert}_{\cU_\zy} 
\equiv \biggl[d \yz \wedge dz  +  d \ybz \wedge d{\bz}\biggr
]
\lbl{omegaz}
\end{eqnarray}

is closed and 

\begin{eqnarray}
\int_{\Sigma} \Omega =\int_{\prt\Sigma}\theta 
\end{eqnarray}
Let us consider now a frame $\cU_\ZY$ : $\theta$   will take the 
form on $T^{*}\ZY$

\begin{eqnarray}
{\theta\arrowvert}_{\cU_\ZY}=\biggl[\YZ dZ   + \YbZ    d\bZ  \biggr]
\lbl{thetay}
\end{eqnarray} 

$\Omega$ is globally defined  and in $\cU_{\ZY}$ 
is defined as:
\begin{eqnarray}
{\Omega\arrowvert}_{\cU_\ZY}=d\theta=\biggl[d\YZ \wedge dZ  +
d\YbZ \wedge  d\bZ \biggr]
\lbl{omegaZ}
\end{eqnarray}

If we require the invariance of the theory under diffeomorphisms,
we have to impose that the local change of frame will generate a
canonical 
transformation. 
 
The change of charts will be canonical if  in $\cU_\zy\cap \cU_\ZY$
\begin{eqnarray}
{\Omega\arrowvert}_{\cU_\zy}={\Omega\arrowvert}_{\cU_\ZY}
\lbl{canonical}
\end{eqnarray}

this would imply:
\begin{eqnarray}
{\theta\arrowvert}_{\cU_\zy}-{\theta\arrowvert}_{\cU_\ZY}= d F
\end{eqnarray}
$F$ is  a function on  $\cU_\zy\cap \cU_\ZY$.
In the  $\zY$ plane we can define the function $\Phi\zY$ as:

\begin{eqnarray}
 d\Phi(z,Y)\equiv d \biggl (F + (\YZ Z+\YbZ \bZ \biggr)
 =\biggl(\yz  dz + \ybz  d\bz\biggr)+
\biggl(d\YZ Z + d\YbZ  \bZ \biggr)
\lbl{Phi}
\end{eqnarray}

The function  $\Phi(z,Y)$ is the generating function for the 
change of charts and it 
is defined (up to a total differential) on the space $\cU_{\zY}$ and has
non vanishing Hessian, ${\dps ||\frac{\prt^2 \Phi}{\prt z \prt Y}||}$.

In the region $\cU_{\zY}$ the differential operator takes the form
\begin{eqnarray}
d=dz \prtz  +d \bz\prtbz   + d\YZ \frac{\prt}{\prt \YZ} +d\YbZ 
\frac{\prt}{\prt \YbZ}\equiv (d_z + d_{\YZ} )
\end{eqnarray}

and $d^2=0$ will imply:

\begin{eqnarray}
\biggl[\prtz,\frac{\prt}{\prt \YZ} \biggr]=0\qquad
\biggl[\prtbz,\frac{\prt}{\prt \YZ} \biggr]=0
\lbl{commzy}
\end{eqnarray}
(and the c.c. commutators).

If we impose $d^2\Phi=0$ we get the important properties:
\begin{eqnarray}
\bprt \yz= \prt\ybz\qquad
\frac{\prt}{\prt \YbZ} Z=\frac{\prt}{\prt \YZ} \bZ 
\lbl{integr1}\nn \\
\frac{\prt}{\prt \YZ}\yz=\prt Z\qquad
\frac{\prt}{\prt \YbZ}\yz=\prt \bZ 
\lbl{integr2}
\end{eqnarray}
and their c.c., which  yield that the mappings:
\begin{eqnarray}
\yz\zY=\prt\Phi(z,Y)
 \lbl{Phi1}, \hskip 1.cm
 Z\zY=\frac{\prt}{\prt \YZ} \Phi(z,Y)
\lbl{Phi2}
\end{eqnarray}
are canonical.
In particular for an arbitrary  surface $\YZ= const$ we can
 construct a local change of coordinates:
\begin{eqnarray}
\zbz\lra\ZBZ\quad for \quad Z\zbz=\frac{\prt}{\prt \YZ}
\Phi(z,Y)\arrowvert_{\YZ=const}
\lbl{diff1}
\end{eqnarray} 

In the following, for writing convenience, we shall choose this constant
equal to zero.
\bigskip
Going on we can verify that:

\begin{eqnarray}
{\theta\arrowvert}_{\cU_\zy}=d_z \Phi(z,Y)
\end{eqnarray}

So ${\Omega\arrowvert}_{\cU_{\zY}}$
takes the elementary form:

\begin{eqnarray}
{\Omega\arrowvert}_{\cU_{\zY}} &=& d{\theta\arrowvert}_{\cU_\zY}
=d\YZ\wedge dz \frac{\prt}{\prt \YZ}\frac{\prt}{\prt
z}\Phi\zY+d\YbZ\wedge dz 
\frac{\prt}{\prt \YbZ}\frac{\prt}{\prt z}\Phi\zY\nn \\
&&+\ d\YZ\wedge d\bz \frac{\prt}{\prt \YZ}\frac{\prt}{\prt \bz}\Phi\zY
+ d\YbZ\wedge d\bz \frac{\prt}{\prt \YbZ}\frac{\prt}{\prt \bz}\Phi\zY\nn
\\
&=& d\YZ\wedge d_z Z +d\YbZ\wedge d_z \bZ=d_{\YZ}
 {\yz} \wedge dz +d_{\YbZ}{\ybz} \wedge d\bz\nn \\
&=& d_Y d_z\Phi\zY
\lbl{omegafull}
\end{eqnarray}

We shall now introduce  some quantities which will be useful for our
treatment.

Let us call

\begin{eqnarray}
\lambda\zY =\prtz\prtY\Phi\zY\qquad
\lambda\zY\mu\zY=\prtbz\prtY\Phi\zY\nn\\
\blambda\zY=\prtbz\prtbY\Phi\zY\qquad
\blambda\zY\mub\zY=\prtz\prtbY\Phi\zY\nn\\
\lbl{parameters}
\end{eqnarray}

the above expression will take the form:

\begin{eqnarray}
{\Omega\arrowvert}_{\cU_\zY}&=& \biggl[dz \wedge\biggl(\lambda 
d\YZ+\blambda\zY\mub\zY d\YbZ\biggr)
+ d \bz  \wedge\biggl(\blambda\zY d\YbZ+\lambda\zY\mu\zY
d\YZ\biggr)\biggr]\nn\\
&=&
 \biggl[\lambda\zY\biggl(dz +\mu\zY d\bz\biggr)\wedge 
d\YZ
+
 \blambda\zY \biggl( d \bz + \mub\zY dz  \biggr)\wedge d\YbZ\biggr]
\end{eqnarray}

due to the global definition of ${\Omega\arrowvert}_{\cU_\zY}$ we can 
derive:

\begin{eqnarray}
d_z Z\zY=\lambda\zY\biggl(dz +\mu\zY d\bz\biggr)\\
d_Y\yz\yZ=\lambda\yZ\biggl( d\YZ+\frac{\blambda\yZ\mub\yZ}{\lambda\yZ}
d\YbZ\biggr)
\lbl{dd}
\end{eqnarray}
(and their c.c.)
which  reveal a complex  structure  parametrized  by 
an  ordinary  Beltrami  multiplier $\mu\zY$ in the $\zbz$ plane 
by ${\dps \frac{\blambda\mub}{\lambda}}$ in the $\YbY$ one.

So the $Z\yZ$ and $y\zY$ coordinate systems can be defined in terms of
a given $\mu\zY$, by means of the Equations:

\begin{eqnarray}
(\bprt -\mu\zY\prt)Z\zY=0 \\
(\prtY -\frac{\lambda\yZ\mub\yZ}{\lambda\yZ}\prtbY) \yz\zY  =0
\end{eqnarray}
 
From the previous equations the Liouville theorem will follow:
\begin{eqnarray}
\det |\frac{\prt Z\yZ}{\prt z}|
=\lambda\yZ\blambda\yZ(1-\mu\yZ\mub\yZ)
 =
\det |\frac{\prt \yz\yZ}{\prt Y}|
\end{eqnarray}

Using  the  previous  parametrization, as is well known, we can write 
the derivative operators $\prtZ,\frac{\prt}{\prt\YZ}$ as:
\begin{eqnarray}
\prtZ=\frac{\prtz-\mu\yZ\prtbz}{\lambda\yZ(1-\mu\yZ\mub\yZ)
} \\
\frac{\prt}{\prt\yz\zY}=
\frac{\cD^z-\mu\zY\cbD^{\bz}}{1-\mu\zY\mub\zY}
\lbl{prty}
\end{eqnarray}
(and their c.c) where we have introduced:
\begin{eqnarray}
\cD^z\zY=\frac{1}{\lambda\zY}\frac{\prt}{\prt\YZ}
\lbl{DY0}
\end{eqnarray}
Finally, if we work in the ${\cU_\zY}$ space, taking $z$ and $\YZ$ as
passive 
coordinates,
the condition $ d{\Omega\arrowvert}_{\cU_\zY} =0 $ will give:
\begin{eqnarray}
d {\Omega\arrowvert}_{\cU_\zY} &=&  \bprt \lambda\yZ d\bz
\wedge dz\wedge d\YZ 
+ \prt(\lambda\yZ\mu\yZ)dz\wedge d\bz\wedge d\YZ \nn \\
&&+\ \prt \blambda\yZ dz\wedge d\bz\wedge d\YbZ 
+ \bprt(\blambda\yZ \mub\yZ) d\bz\wedge dz\wedge d\YbZ  \nn\\
&& +\ d\YbZ\wedge \frac{\prt}{\prt \YbZ}
\biggl( \lambda\yZ\biggl(dz +\mu\yZ d\bz\biggr)\biggr)\wedge d\YZ\nn\\
&&+\ d\YZ\wedge \frac{\prt}{\prt \YZ}
\biggl( \blambda\yZ\biggl(d\bz +\mub\yZ dz\biggr)\biggr)\wedge d\YbZ =0
\end{eqnarray}
which gives rise to the Beltrami identities:
\begin{eqnarray}
\bprt \lambda\zY =\prt(\mu\zY \lambda\zY), \qquad
\frac{\prt}{\prt \YbZ} \lambda\zY = \frac{\prt}{\prt \YZ} \biggl
(\blambda\zY \mub\zY\biggr)
\lbl{beltra}
\end{eqnarray}
and their c.c.

It is important to remark that from Eq \rf{beltra} one has,
\begin{eqnarray}
\frac{\cD^z \blambda\yZ}{\blambda\yZ}=\frac{\cbD^\bz \mu\yZ
+\mu\yZ\cD^z\mub\yZ}{1-\mu\yZ\mub\yZ}
\lbl{beltraD}
\end{eqnarray}
 and its c.c. From Eq \rf{commzy} we also get:
\begin{eqnarray}
\biggl[\prtz,\cD^z\biggr]&=&-\prtz\log\lambda\yZ\cD^z\nn\\
\biggl[\prtbz,\cD^z\biggr]&=&-\prtbz\log\lambda\yZ\cD^z
=-\biggl(\prt\mu\yZ\cD^z-\mu\yZ\biggl[\prtz,\cD^z\biggr]\biggr)
\end{eqnarray}
(and their c.c.), and the commutator,
\begin{eqnarray}
\biggl[\cD^z,\cbD^\bz\biggr]=(\cD^z\mub\yZ)\frac{\prt}{\prt\yz\yZ}-(\cbD^\bz\mu\yZ)\frac{\prt}{\prt\ybz\yZ}
\end{eqnarray}
from which it follows:
\begin{eqnarray}
\biggl[\frac{\prt}{\prt\yz\yZ},\frac{\prt}{\prt\ybz\yZ}\biggr]=0.
\end{eqnarray}
At this stage, one remark is in order. In Eq \rf{omegafull}
the terms $d_{\YZ} Z \wedge d{\YZ}  +d_{\YZ}\bZ\wedge d{\YbZ} $ 
and $ d_{z}\yz \wedge dz + d_z \ybz \wedge d\bz $ will identically
vanish in $\Omega$. 
Accordingly, we can state the important Theorem \cite{BaLa0}:  
\begin{guess}
On the smooth trivial bundle $\Sigma\times\mbox{\bf R}^2$,
the vertical holomorphic change of local coordinates, 
\begin{eqnarray}
Z(\zbz ,\YbY) \lra Z(\zbz, \cF(Y_Z),\YbZ),
\end{eqnarray}
where $\cF$ is a holomorphic function in $\YZ$, while the horizontal
holomorphic change of local coordinates,
\begin{eqnarray} 
\yz(z,\bz,\YbY)\lra \yz(f(z),\bz,\YbY),
\end{eqnarray}
where $f$ is a holomorphic function in $z$,
are both canonical transformations.
 \end{guess}
In the first case 
an infinitesimal variation of $Z\zY$ in $\YZ$ does not modify, for fixed
$\zbz$ the $\Omega$ form.  

So the diffeomorphisms $\z \lra Z(\zbz,\YZ);
z \lra Z(\zbz,\YZ +d\YZ)$,
will be related to the same two form $\Omega$.
 
If we make the expansion around, say,$\YZ=0,\YbZ=0$  the 
generating function $\Phi$  will be written as the series :
\begin{eqnarray}
\Phi\zY &=& \sum_{n=1}^{n_{max}}\frac{1}{n!}
\biggl[\YZ^n{\biggl(\frac{\prt}{\prt \YZ}\Phi_1(\zbz,\YZ)\biggr)}^n
\arrowvert_{\YZ=0,\YbZ=0}\biggr]\nn \\
&&\hskip .5cm +\ \sum_{n=1}^{n_{max}}\frac{1}{n!}
\biggl[\YbZ^n{\biggl(\frac{\prt}{\prt
\YbZ}\bar{\Phi_1}(\zbz,\YbZ)\biggr)}^n
\arrowvert_{\YZ=0,\YbZ=0}\biggr]\nn \\
&\equiv&
\sum_{n=1}^{n_{max}}\biggl[\YZ^n\, \Zn\zbz\biggr]+\sum_n\biggl
[\YbZ^n\, \Zbn\zbz\biggr]
\lbl{phi}
\end{eqnarray}
 where:
\begin{eqnarray}
\Zn\zbz\equiv\biggl[\frac{1}{n!}{\biggl(\frac{\prt}{\prt \YZ}\biggr)}^n
\Phi_1\zY\biggr]
\arrowvert_{\YZ=0,\YbZ=0}
\end{eqnarray}
 
And we can reconstruct $Z\zY$ as:
\begin{eqnarray}
Z\zY= \sum_n n Z^{(n)}\zbz {\YZ}^{n-1}
\end{eqnarray}
 
We shall see in the following that the family of reparametrizations:

\begin{eqnarray}
\zbz \lra \ZBZn
\lbl{wtransformations}
\end {eqnarray}
 will be the origin of the $w$ 
algebras symmetry transformations.
Obviously the choice of the $\YbY$ origin as
 starting point does not alter the treatment

 The symplectic form will then be written: 
\begin{eqnarray}
{\Omega\arrowvert}_{\cU_{\zY}}
=d_Y d_z\Phi\zY=\sum_n\biggl[d_Y\YZ^n\wedge d_z \Zn\zbz
 + d_Y\YbZ^n \wedge d_z \Zbn\zbz\biggr]
\lbl{omegafull1}
\end{eqnarray}

Note that the conjugate momenta to
 the complex coordinates $Z^{(n)},\ \forall n\neq 1$ are 
 related to the n-th power of the conjugate 
 momenta of the coordinate $Z^{(1)}\zbz$.
 
\bigskip 
 
\sect{The geometry of the $\Zn$-spaces} 

\indent

In the previous chapter we introduced  
 the mappings:
\begin{eqnarray}
\zbz \lra \ZBZn\quad\forall n=1\cdots n
\end{eqnarray}
foreseeing that they 
will be fundamental for our purposes.

Before investigating their role in the construction of $w$-algebras,
we shall derive down below the most important properties of the
$\ZBZn$-spaces.  

\subsect{Generalities on $\ZBZn$-spaces}
 
 \indent

The $\Zn\zbz$ coordinates are defined as:
 \begin{eqnarray}
\leqa{\Zn\zbz=\biggr[
\frac{1}{n!} {\biggl(\frac{\prt}{\prt \YZ}\biggr)}^{n}
(\Phi\zY)\biggr]\arrowvert_{\YZ,\YbZ=0}} \nn\\
&=& \sum_{j=1,\cdots ,n} j! \prod_{i=1,\cdots, m_j}\biggl[
\frac{{(\prt Z^{(p_i)}\zbz)}^{a_i}}{{a_i}!}
\biggr]\arrowvert_
{\tiny{\left\{
\begin{array}{c}
{\sum_i a_i =j}\\
{\sum_i a_i p_i=n}\\
{p_1>p_2>,\cdots,>p_{m_j}} 
\end{array}\right\}}}
\cM^{(j)}\zbz 
\lbl{Zdecomp}
\end{eqnarray}

where the functions $\cM^j\zbz$ are given by:
\begin{eqnarray} 
\cM^{(j)}\zbz\equiv\frac{1}{j!}{\biggl(\cD\biggr)}^j\Phi\zY 
\arrowvert_{\YZ,\bar{\YZ}=0} 
\end{eqnarray}

Note that the $\Zn\zbz$ coordinate is no
 more independent from  $\Zr$,
for $r=1,\cdots ,n-1$; by the way it obeys differential consistency
conditions 
induced by the $\cM^{(j)}\zbz$ functions:
we shall sketch the above system for further convenience:
\begin{eqnarray}
 Z\zbz  & = & \prt Z\cM^{(1)}\zbz\nn \\
 Z^{(2)}\zbz & = & \prt Z^{(2)}\zbz\cM^{(1)}\zbz +{{(\prt Z)}^2}
 \cM^{(2)}\zbz\nn \\              
Z^{(3)}\zbz & =  & \prt Z^{(3)}\zbz\cM^{(1)}\zbz
 +2\prt Z\zbz \prt Z^{(2)}\zbz\cM^{(2)}
 \zbz +{{\prt Z\zbz}^3}\cM^{(3)}\zbz\nn \\
\vdots \nn\\ 
Z^{(N)}\zbz  & = & \prt Z^{(N)}\zbz  \cM^{(1)}
\zbz 
+\cdots\cdots\cdots +{\biggl(\prt Z\zbz\biggr)}^N \cM^{(N)}\zbz 
\lbl{system}
\end{eqnarray}
where the first equation can be solved by
\begin{eqnarray}
\ln Z\zbz =\int^z_{\tilde{z}} dz' \frac{1}{\cM^1(z',\bz)}+
\ln Z(\tilde{z},\bz)
\end{eqnarray}
where $\ln Z(\tilde{z},\bz)$ takes into account the boundary conditions.
Thus, one has
\begin{eqnarray}
Z\zbz=Z(\tilde{z},\bz)\exp{\int^z_{\tilde{z}} dz'
\frac{1}{\cM^{(1)}(z',\bz)}}
\end{eqnarray}
which plugged into the second equation  gives 
$Z^{(2)}$  as an integral relation between $Z$,
 $\cM^{(1)}\zbz$ and $\cM^{(2)}\zbz$;
 \bigskip
\begin{eqnarray} 
Z^{(2)}\zbz=Z\zbz
\biggl[
\int^z dz"
 \frac{\cM^{(2)}(z",\bz)Z(z",\bz)}{{(\cM^1(z",\bz))}^3}\biggr] 
 \end{eqnarray}
In full generality, one gets,
\begin{eqnarray} 
Z^{(N)}\zbz=Z\zbz
\biggl[
\int^z dz"\frac{1}{Z(z",\bz)} 
 {\cal F}^{N}(Z^{(j)}({\cM^k(z",\bz)}_{k\leq j})\cM^i(z",\bz))\biggr]
 \end{eqnarray}
where $\cM^{(i)}\zbz\quad i=1,2,\cdots, n$ are fixed and
$Z^{(j)}(z",\bz)\quad j<n $
 have been  calculated at the preceding  orders. So we can state the
following 
  
\begin{guess}
 The set of functions $\cM^{(i)}\zbz$, $i=1,\cdots, n$ completely 
 identify the set of coordinates $Z^{(j)}\zbz$, $j=1,\cdots ,n$ 
\end{guess}  
 
 Now we want  to solve another problem: 
 given a 
local change of coordinates $\zbz \lra (\cZ\zbz ,\bar{\cZ}\zbz )$
  is it  possible to consider it as  an element of arbitrary n-th
   order of a $w$ hierarchy  $\zbz\lra \ZBZn$, and to find a construction
of the underlying $(Z^{(i)},\bZ^{(i)})\quad i=1,\cdots, n-1$ in order
 to get a $w$ description of this local change?  
 
The answer is positive, since using   a standard  construction of
 the  $(Z^{(i)},\bZ^{(i)}),\ i=1,\cdots ,n-1$ spaces (which do
 not interfere with  $\ZBZn$),  we can use it in the last equation
 of \rf{system} and  get in an algebraic way the suitable
 solution $\cM^N\zbz$. So we can state the Theorem:

\begin{guess}
For an arbitrary local change:
 $\zbz \lra (\cZ\zbz ,\bar{\cZ}\zbz )$ 
it is possible to generate a space hierarchy
 $\zbz\lra \ZBZr,\ r=1,\cdots, n$. 
 \lbl{guess3} 
\end{guess}  

Finally we explore  the $\ZBZn$ spaces  with respect to the $\zbz$
background.
 As in \cite{BaLa}, we shall introduce: 

\begin{eqnarray}
d_z \Zn\zbz &=& \biggl
[\frac{1}{n!} {\biggl(\frac{\prt}{\prt \YZ}\biggr)}^n\prt\Phi\zY
dz +\frac{1}{n!} {\biggl(\frac{\prt}{\prt \YZ}\biggr)}^n\bprt\Phi\zY
d\bz\biggr]
\arrowvert_{\YZ=0,\YbZ=0}\nn \\
&\equiv& \lambda_z^{\Zn}\zbz [dz +\mu_\bz^z(n,\zbz) d\bz] 
\lbl{beltran0}
\end{eqnarray}

where:
\begin{eqnarray}
\lambdan\zbz\equiv \biggl[\frac{1}{n!} {\biggl(\frac{\prt}{\prt \YZ}
\biggr)}^n\prt\Phi\zY\biggr]\arrowvert_{\YZ=0,\YbZ=0}\equiv \prt \Zn\zbz
\nn \\
\lambdan\zbz \mu_\bz^z(n,\zbz) \equiv
\biggl[\frac{1}{n!} {\biggl(\frac{\prt}{\prt
\YZ}\biggr)}^n\bprt\Phi\zY\biggr]
\arrowvert_{\YZ=0,\YbZ=0}
\equiv \bprt \Zn\zbz.
\lbl{lambdan}
\end{eqnarray}
So we shall define, for all $n$:
\begin{eqnarray}
\frac{\prt}{\prt \Zn} =\frac{\prt -\mub (n,\zbz)
\bprt}{(1-\mu(n,\zbz)\mub(n,\zbz))}
\end{eqnarray}
In particular we get from Eq \rf{Zdecomp}:
\begin{eqnarray}
\frac{\prt}{\prt \Zbn}\biggl( \sum_{j=1}^n j! \prod_{i=1}^{m_j}\biggl[ 
\frac{ {(\lambdapi\zbz))}^{a_i} }{{a_i}!}
\biggr]\arrowvert_
{\tiny{\left\{
\begin{array}{c}
{\sum_i a_i =j}\\
{\sum_i a_i p_i=n}\\
{p_1>p_2>,\cdots,>p_{m_j}} 
\end{array}\right\}}}
\cM^{(j)}\zbz\biggr)=0.
\lbl{Zderivat}
\end{eqnarray}
This identity which now appears trivial, will acquire a particular
meaning in the following.  
   
It is so evident that the quantity $\mu_\bz^z(n,\zbz)$ will label the 
complex structure of the space $\Zn$ in the $\zbz$ background
and increasing the order $n$ this complex structure will explore the
complex
structure of all the $\zY$ space.

More precisely we get:
\begin{eqnarray}
\lambda\zY \mu\zY &=& \sum_n n \lambdan\zbz \mu(n,\zbz) {\YZ}^{n-1}\nn
\\
\lambda\zY &=& \sum_n n \lambdan \zbz {\YZ}^{n-1}
\end{eqnarray}

The symplectic form will reduce to:
  \begin{eqnarray}
\Omega\arrowvert_{\cU\zY} =\sum_{n=1\cdots n_{max}}\biggl[ \lambdan\zbz
d {\YZ}^n +
 \blambdan \mub (n,\zbz)d{\YbZ}^n \biggr] \wedge dz\nn \\
+\sum_{n=1 \cdots n_{max}}\biggl[ \blambdan\zbz d{\YbZ}^n +
  \lambdan\zbz \mu(n,\zbz)d{\YZ}^n\biggr]\wedge d\bz 
\end{eqnarray}

 From the very definition, the Beltrami parameter  will take 
the general form:

\begin{eqnarray}
 \mu_\bz^z(n,\zbz)\equiv \frac{\bprt Z^{(n)}\zbz}{\prt Z^{(n)}\zbz} 
=\sum_{j=1}^n j! \prod_{i=1}^{m_j}\biggl[
\frac{{(\lambdapi\zbz)}^{a_i}}{{a_i}!\lambdan\zbz}
\biggr]\arrowvert_
{\tiny{\left\{
\begin{array}{c}
{\sum_i a_i =j}\\
{\sum_i a_i p_i=n}\\
{p_1>p_2>,\cdots,>p_{m_j}} 
\end{array}\right\}}}
\mu^{(j)}_\bz\zbz.
\lbl{mun}
\end{eqnarray}
where we have introduced:
\begin{eqnarray}
\mu_{\bz}^{(n)}\zbz=\biggl[
\frac{1}{(n)!}{\biggl(\cD^z\biggr)}^{n}\bprt\Phi\zY\biggr]
\arrowvert_{\YZ,\YbZ=0}
\lbl{mun0}
\end{eqnarray}

We remark that, due to Eq.\rf{dd} the presence of $\lambda$'s in the $\YZ$
derivative compromises the locality requirements but the parameters 
in Eq \rf{mun0} introduce a  suitable parametrization  
for a local Lagrangian Quantum Field Theory use.
Furthermore these ones have to be considered as the least common factors
for all the Beltrami factors  of the spaces $\Zr\zbz\quad r=1,\cdots ,n$

Note that  the Beltrami multiplier $\mu_\bz^z(n,\zbz)$ will depend
on the $\lambda 's$
parameters of the spaces parametrized by the $Z^{(i)}$ coordinates
with $i\leq n$. 

Under a change of background coordinates the Beltrami multiplier 
transforms as:

\begin{eqnarray}
\mu_\bz^z(n,\zbz)= 
\mu^{z'}_{\bz'}(n,\zbzp)(\prt' z)(\bprt \bz')
\end{eqnarray}
so we can derive: 
\begin{eqnarray}
\mu_\bz^{(n)}\zbz= 
\mu^{(n')}_{\bz'}\zbzp{(\prt' z)}^n(\bprt \bz')
\end{eqnarray}
 giving a well-defined geometrical status to $\mu_\bz^n$ as a
 $(-n,1)$-conformal field.  

A Beltrami identity is immediately recovered for each (and any) $n$ as
a consequence of $d_z^2=0$,

\begin{eqnarray}
\biggl[\frac{1}{n!} {\biggl(\frac{\prt}
{\prt \YZ}\biggr)}^n\prt\bprt\Phi\zY
\biggr]
\arrowvert_{\YZ=0,\YbZ=0}
=
\bprt\lambdan\zbz= \prt(\lambdan\zbz\mu_\bz^z(n,\zbz))
\lbl{beltramin}
\end{eqnarray}

It is not only obvious that $\lambdan$ is a non local object   on
 $\mu_\bz^z(n,\zbz)$, 
but, due to Eq.\rf{mun0} the parameters 
$\lambdan$ is not local on the parameters $\mu^{(i)}_\bz\zbz$ 
 with $i\leq n$. Furthermore from Eq \rf{mun} we can realize that 
the Beltrami multiplier $\mu_\bz^z(n,\zbz)$ (see  
Eqs\rf{beltramin} \rf{mun0} ) is sensible to the 
complex structures of the inner subspaces ; so we can state:  
 
\begin{statement}
\item{a) The  $\lambdan$ (for fixed n) are non local functions of
the parameters $\mu_{\bz}^{(r)}$, and will contain all of them with order
$r\leq n$}.

\item{b)The complex structure of the $\ZBZn$ spaces  can be described by 
parameters $\mu_\bz^z(n,\zbz)$ which extend to these spaces the  
Beltrami multipliers. For a given $n$,  $\mu_\bz^z(n,\zbz)$ 
will depend, through $\lambda^{\Zr},\quad r<n $, in a  non local
way on  the $\mu_\bz^{(j)}$'s with $j \leq n$.}

\item{c) As a consequence of the previous statements and of
Eq\rf{beltran0}
the coordinate  $\Zn\zbz$ is a non local function of 
 $\mu_\bz^{(j)}\zbz$ with $j\leq n$}
 \lbl{guessZ} 
\end{statement}
These are important geometrical statements of the work 
and are the basis for the 
physical discussion of the problem.
So the mappings

\begin{eqnarray}
\zbz \lra \ZBZn\quad \forall n
\end{eqnarray}
are non holomorphic and the non holomorphicity  depends on $n$.

The only possibility to get local Beltrami multipliers 
$\mu_\bz^z(n,\zbz)$ is to have   
$\mu^{(r)}_\bz\zbz =0\quad \forall r>1$ .
In this case the complex structure of all the $\ZBZn$ space 
will coincide with the $(Z^{(1)},\bZ^{(1)})$ one.
So necessarily $\forall r \geq 2$:
\begin{eqnarray}
{\biggl[\cD^z\biggr]}^r\bprt\Phi\zY\arrowvert_{\YZ,\YbZ=0} =
{\biggl[\frac{1}{\lambda\zY}\frac{\prt}{\prt\YZ}\biggr]}^r\sum_n
\biggl[\YZ^n\bprt \Zn\zbz\biggr]\arrowvert_{\YZ,\YbZ=0}=0,
\end{eqnarray}

It is easy to see that the previous equation leads to:
\begin{eqnarray}
\frac{\bprt Z^{(r)}\zbz}{\prt Z^{(r)}\zbz}
= 
  \frac{\bprt Z\zbz}{\prt Z\zbz } \equiv \mu\zbz,\qquad
\forall r \geq 2
\end{eqnarray}

which, expressed in terms of the $\ZBZ$ background,  gives:
\begin{eqnarray}
(\bprt-\mu\zbz\prt) Z^{(r)}=0 ,\qquad
\forall r \geq 2 ,
\end{eqnarray}
\bigskip
showing that $Z^{(r)}$ is an holomorphic function of $Z$.

This result is straightforwardly generalized.  Imposing $\forall r \geq l+1$: 
\begin{eqnarray}
{\biggl[\cD^z\biggr]}^r\bprt\Phi\zY\arrowvert_{\YZ,\YbZ=0} = 
{\biggl[\frac{1}{\lambda\zY}\frac{\prt}{\prt\YZ}\biggr]}^r\sum_n\biggl[\YZ^n\bprt
\Zn\zbz\biggr]\arrowvert_{\YZ,\YbZ=0}=0,
\end{eqnarray}
one recovers
\begin{eqnarray}
\frac{\bprt Z^{(r)}\zbz}{\prt Z^{(r)}\zbz}= 
  \frac{\bprt Z^{(l)}\zbz}{\prt Z^{(l)}\zbz } \equiv \mu(l,\zbz),\qquad
\forall r \geq l+1
\end{eqnarray}
which leads to:
 \begin{eqnarray}
\bprt_{Z^{(l)}} Z^{(r)}=0,   \qquad \forall r \geq l+1.
\end{eqnarray} 
Thus $Z^{(r)}$ is holomorphic in $Z^{(l)}$.
Obviously the inverse is always true ; so we can state:

\begin{guess}
 The set of conditions:
$\mu_\bz^{(r)}=0\ , r=l+1,\cdots, n$
will imply that $Z^{(n)}$ is an holomorphic function of $Z^{(l)}$ and
vice-versa.  
\end{guess}

\subsect{Complex structures in $\ZBZr$ backgrounds}

It is already interesting
 to analyze the complex structure of $\ZBZn$-space with 
respect to different backgrounds; indeed setting
\begin{eqnarray}
d\Zn =\Lambda^{\Zn}_{Z^{(r)}}\ZBZr
\biggl[d\Zr +\Xi_{\Zbr}^{Z^{(r)}}(n,\ZBZr)d\Zbr\biggr]
\end{eqnarray}
where: 
\begin{eqnarray}
\Lambda^{\Zn}_{Z^{(r)}}\ZBZr\equiv\frac{\prt \Zn\ZBZr}{\prt \Zr}\nn\\
\Lambda^{\Zn}_{Z^{(r)}}\ZBZr\,\Xi_{\Zbr}^{Z^{(r)}}(n,\ZBZr
\equiv\frac{\prt \Zbn\ZBZr}{\prt \Zr}
\end{eqnarray}
so that the quantity $\Xi_{\Zbr}^{Z^{(r)}}(n,\ZBZr)$ is the Beltrami
multiplier of the  
coordinates $\Zn$ in the $\ZBZr$ background $\forall n$. 
So we can relate these quantities to the corresponding objects
relatively to the
$\zbz$ background,  $\forall \ n$ and $r$: 
\begin{eqnarray} 
\Lambda^{\Zn}_{Z^{(r)}}\ZBZr\arrowvert_{\zbz} 
 =
\frac{\lambda_z^{\Zr}\zbz\biggl(1-\mu(r,\zbz)\mub(n,\zbz)\biggr)}
{\lambdan\zbz \biggl(1-\mu(n,\zbz)\mub(n,\zbz)\biggr)}
\mub(n,\zbz)\biggr)
\lbl{lambdar}
\end{eqnarray}

and:

\begin{eqnarray}
\Xi_{\Zbr}^{Z^{(r)}}(n,\ZBZr)\arrowvert_{\zbz}=
\frac{\lambda^{Z^{(r)}}\zbz \biggl(\mu(r,\zbz)-\mu(n,\zbz)\biggr)}
{\blambda^{\Zbr}\zbz\biggl(1-\mu(r,\zbz)\mub(n,\zbz)\biggr)}
\lbl{xir} 
\end{eqnarray}

also we derive:
\begin{eqnarray}
\biggl[\Lambda^{\Zn}_{Z^{(r)}}\ZBZr
\bar{\Lambda}^{(\bZ^{n})}_{(\bZ^{r})}\ZBZr
(1-\Xi_{\Zbr}^{Z^{(r)}}(n,\ZBZr) \bar{\Xi}^{\Zbr}_{Z^{(r)}}
(n,\ZBZr) \biggr]\arrowvert_{\zbz}\nn \\
=\frac{\lambdan\zbz \blambdan\zbz(1-\mu(n,\zbz)\mub(n,\zbz))} 
{\lambda^{Z^{(r)}}\zbz \blambda^{\Zbr}\zbz(1-\mu(r,\zbz)\mub(r,\zbz))} 
\end{eqnarray}

\sect{$w$ B.R.S algebras}

The previous construction introduces to a BRS derivation of
 $w$-symmetry as shown in \cite{BaLa0}. 
 Our aim is to construct a BRS differential which considers, 
 in an infinitesimal approach {\it {all}} the mappings
  $\zbz \lra \ZBZn$, for all $n$, on the same footing.  

Consider first the infinitesimal variations $\Lambda\zY$ of 
the generating function $\Phi\zY$ under the diffeomorphism action of the
 cotangent bundle.
Then by taking the expansion \rf{phi} one can proceed as follows \cite{BaLa0}.  
Let $\sS$ be  the BRS  diffeomorphism operator acting on the $\zbz$
basis and defined for each $n$ by

 \begin{eqnarray}
\sS \Zn\zbz\equiv\Upsilon^{(n)}\zbz\equiv\biggl[\frac{1}{n!}
{\biggl(\frac{\prt}{\prt
\YZ}\biggr)}^n\Lambda\zY\biggr]\arrowvert_{\YZ,\YbZ=0}
\end{eqnarray}

consequently 
the $\sS$ nilpotency will give:

\begin{eqnarray}
\sS\Upsilon^{(n)}\zbz=0
\end{eqnarray}

  we shall decompose:
\begin{eqnarray}
\Upsilon^{(n)}\zbz\equiv\biggl[\lambdan\zbz \cK_n^z\zbz\biggr]
\end{eqnarray} 

and we get from the definition of $\lambdan\zbz$ :
\begin{eqnarray}
\sS\lambdan\zbz &=& \biggl[\frac{1}{n!} {\biggl(\frac{\prt}{\prt \YZ}
\biggr)}^n\prt\Lambda\zY\biggr]\arrowvert_{\YZ,\YbZ=0}\nn\\ 
 &\equiv& \biggl[\prt\biggl(
\frac{1}{n!} {\biggl(\frac{\prt}{\prt \YZ}
\biggr)}^{n-1}(\lambda\zY\cC\zY)\biggr)\biggr]\arrowvert_{\YZ,\YbZ=0}
\nn \\
&=& \prt\biggl(\lambdan\zbz \cK_n^z\zbz\biggr)
\lbl{slambdan}
\end{eqnarray}

and consequently: 
\begin{eqnarray}
\sS\cK_n^z\zbz=\cK_n^z\zbz\prt\cK_n^z\zbz
\lbl{skn}
\end{eqnarray}

Expanding the calculations yields: 
\begin{eqnarray}
\cK_n^z\zbz 
=\sum_{j=1}^n j! \prod_{i=1}{m_j}\biggl[
\frac{{(\lambdapi\zbz)}^{a_i}}{{a_i}!\lambdan}
\biggr]\arrowvert_
{\tiny{\left\{
\begin{array}{c}
{\sum_i a_i =j}\\
{\sum_i a_i p_i=n}\\
{p_1>p_2>\cdots>p_{m_j}} 
\end{array}\right\}}}
\cC^{(j)}\zbz
\lbl{kn}
\end{eqnarray}
where we have introduced:
\begin{eqnarray}
\cC^{(n)}\zbz 
=\biggl[\frac{1}{n!} {\biggl(\cD^z\biggr)}^n \Lambda\zY
\biggr]\arrowvert_{
{\YZ,\YbZ=0}},\qquad
n=1,2\cdots 
\lbl{0000000b}
\end{eqnarray}
which, for all $n$, provide, in Eq \rf{kn}, a geometric expansion
 with the same non local coefficients as in  Eq \rf{mun}.

It is quite easy, from the very definition, to derive that 
these ghosts transform as:
\begin{eqnarray}
\sS \cC^{(n)}=\sum_{r=1}^n \biggl(r\,\cC^{(r)}\prtz
\cC^{(n-r+1)}  \biggr)
\lbl{000b}
\end{eqnarray}
revealing the holomorphic $w$ character of these ghosts.  We
remark that the B.R.S. variations of $\cC^{(n)}\zbz$ depends  on
the fields $\cC^{(r)}\zbz$ with $r\leq n$.  The upper limit of the
indices of these ghosts coincides with the the upper index of the
expansion Eq \rf{phi},and will characterize this symmetry: we do not
fix it, and our  conclusions hold their validity for any value
(finite or infinite) of this index.

The connection between Eqs \rf{kn} and \rf{mun} spreads new light
 on the connection between diffeomorphisms and $w$ algebras by
 putting into the game the  complex structures.

Now the coefficients of these expansions are essentially
geometrical factors.

On the other hand the quantities $\mu_\Zbn\zbz$ in Eq \rf{mun0}
will have the B.R.S variations:

\begin{eqnarray}
\sS \mu_{\bz}^{(n)}\zbz=\bprt C^{(n)}\zbz -\sum_{r=1}^n
\biggl[ r\, \mu_{\bz}^{(r)}\zbz \prt C^{(n-r+1)}\zbz \nn \\
-\ r\, C^{(r)}\zbz \prt
\mu_{\bz}^{(n-r+1)}\zbz\biggr]
\lbl{smun1}
\end{eqnarray}

and: 
\begin{eqnarray}
\sS\biggl[ \lambdan\zbz \mu_\bz^z(n,\zbz)\biggr] = \bprt \Upsilon^{(p)}
\end{eqnarray}

so that:
\begin{eqnarray}
\sS  \mu_\bz^z(n,\zbz) = \cK_n^z\zbz\prt\mu_\bz^z(n,\zbz)
-\mu_\bz^z(n,\zbz)\prt
\cK_n^z\zbz+\bprt\cK_n^z\zbz
\end{eqnarray}

So for each $n$ a diffeomorphic structure with a ghost $\cK_n^z$ 
(non local in the complex structure parameter )
 can be put into evidence
 \bigskip 
From Eqs \rf{mun} it is easy to find the form of these variations 
in terms of $w$ components.
We shall introduce:

\begin{eqnarray}
\kappa_n^z\zbz\equiv
\frac{\cK_n^z\zbz-\mu(n,\zbz)\bar{\cK_n}^{\bz}\zbz}
{1-\mu(n,\zbz)\mub(n,\zbz)}
\lbl{kappa} 
\end{eqnarray}

for which:

\begin{eqnarray}
\Upsilon^{(n)}\ZBZn\prt_{\Zn}
+\bar{\Upsilon}^{(\bar{n})}\ZBZn\prt_{\Zbn}=
\kappa^z_n\zbz\prt +\bar{\kappa^z_n}\zbz\bprt
\end{eqnarray}

so:
\begin{eqnarray}
\cK_n^z\zbz =\kappa_n\zbz +\mu(n,\zbz)\bar{\kappa}_n^\bz\zbz
\end{eqnarray}
In this base we have:

\begin{eqnarray}
\sS\kappa_n^z\zbz=\kappa_n^z\zbz\prt\kappa_n^z\zbz+\bar{\kappa_n}^\bz\zbz
\bprt\kappa_n^z\zbz
\lbl{skappa}
\end{eqnarray}

\begin{eqnarray}
\sS\mu(n,\zbz)&=&(\kappa_n^z\prt+\bar{\kappa_n}^\bz\bprt)\mu(n,\zbz)
-\mu(n,\zbz)(\prt\kappa_n^z+\mu(n,\zbz)\prt\bar{\kappa}^\bz)\nn\\
&&+\ \bprt\kappa_n^z+\mu(n,\zbz)\bprt\bar{\kappa_n}^\bz
\lbl{smuk} 
\end{eqnarray}
 
\begin{eqnarray}
\sS\lambdan\zbz=(\kappa_n^z\prt+\bar{\kappa_n}^\bz\bprt)\lambdan\zbz
+\lambdan\zbz(\prt\kappa_n^z+\mu(n,\zbz)\prt\bar{\kappa}^\bz)
\lbl{slambdank}
\end{eqnarray}

Note that the condition:
 \begin{eqnarray}
 \sS\Upsilon^{(n)}\zbz =0
 \end{eqnarray}
  is verified only if the Beltrami condition Eq\rf{beltramin} holds.   

In this approach we can also derive:
\begin{eqnarray}
\cC^{(j)}\zbz = 
\sum_{\scriptsize \begin{array}{c} r,s=0\\ r+s>0\end{array}}^j
\biggl[r!s! \biggl(\Pi_{i}\frac{\biggl({\mu^{(l_i)}_\bz\zbz
\biggr)}^{k_i}}{k_i!}\biggr)\arrowvert
_{\tiny{\left\{\begin{array}{c}
{\sum_i k_i =s}\\
{r+\sum_i l_i k_i =j} 
\end{array}\right\}}}\biggr] c^{(r,s)}\zbz
\lbl{Cc1} 
\end{eqnarray}

where we have introduced, in the spirit of Eq \rf{prty} the new ghosts:
\begin{eqnarray}
c^{(p,q)}\zbz 
=\biggl[\frac{1}{p!}\frac{1}{q!} {\biggl(\frac{\prt}{\prt \yz\zY
}\biggr)}^p {\biggl(\frac{\prt}{\prt \ybz\zY}\biggr)}^q
\Lambda\zY\biggr]\arrowvert_{\YZ,\YbZ=0}\nn\\
\lbl{0000000a}
\end{eqnarray}
of conformal weight $(-p,-q)$ and which transform as:

\begin{eqnarray}
\sS c^{(p,q)}\zbz= \sum_{\scriptsize{ \begin{array}{c} r,s=0\\ r+s\geq
1\end{array}}}^{r=p,s=q} 
\biggl(rc^{(r,s)}\zbz\prtz c^{(p-r+1,q-s)}\zbz  \nn \\
+ s c^{(r,s)}\zbz\prtbz
c^{(p-r,q-s+1)}\zbz\biggr)
\lbl{0000a}
\end{eqnarray}

We also remark that the variation  of $c^{(p,q)}$  contains 
the ghost fields $c^{(r,s)}$ with lower degrees, $r\leq p;s\leq q$.

Combining together Eqs.\rf{0000000b} and \rf{Cc1}  we shall write:
\begin{eqnarray}
\cK_n^z\zbz = \sum_{\scriptsize{ \begin{array}{c} r,s=0\\ r+s\geq
1\end{array}}}^n \eta^z_{(r,s)}(n,\zbz)c^{(r,s)}\zbz
\lbl{cA} 
\end{eqnarray}
where
 \begin{eqnarray}
\leqa{ \eta^z_{(r,s)}(n,\zbz)  \
= \sum_{j=\mbox{\scriptsize max}(r,s)}^n  j!
\prod_{i=1}^{m_j} \biggl[ \frac{{(\lambdapi\zbz)}^{a_i}}{{a_i}!\lambdan} 
\biggr] } \\
&\times& \biggl[r!s! \biggl(\Pi_{i}\frac{\biggl({\mu^{(l_i)}_\bz\zbz
\biggr)}^{k_i}}{k_i!}\biggr)\biggr]\arrowvert
_{\tiny{\left\{
\begin{array}{c}
{\sum_i k_i =s}\\
{r+\sum_i l_i k_i =j}\\
{\sum_i a_i =j}\\  
{\sum_i a_i p_i=n}\\
{p_1>p_2>\cdots>p_{m_j}} 
\end{array}\right\}}}\nn
\lbl{eta}
\end{eqnarray}

in particular:
\begin{eqnarray}
\eta^z_{(1,0)}(n,\zbz) &=& 1\nn \\ 
\eta^z_{(0,1)}(n,\zbz) &=& \mu_\bz^z(n,\zbz) 
\end {eqnarray}

The same can be written in terms of the $\kappa_n\zbz$ ghosts, getting:
\begin{eqnarray}
\leqa{\kappa^z_n\zbz\ =\ c^{(1,0)}\zbz} \nn\\
&&+\sum_{\scriptsize{ \begin{array}{c} r,s=0\\ r+s\geq 2\end{array}}}^n 
\frac{\biggl(\eta^z_{(r,s)}(n,\zbz) 
-\mu(n,\zbz)\bar{\eta}^\bz_{(r,s)}\zbz\biggr)}{1-\mu(n,\zbz)\mub(n,\zbz)}
c^{(r,s)}\zbz.
\lbl{kappaA}
\end{eqnarray}

\subsect{The introduction of matter field sectors in $w_n$-algebras}

The introduction of matter in $w$ invariant models, and in particular
$w$ gravity, 
have been treated in the literature in different scenarios according to
the different 
attempts to reach $w$ algebras, e.g \cite{Hull}.

Our point of view, which relates in a geometrical fashion 
$w$ algebras to two dimensional conformal field theory, heavily supports
the 
methods which introduce  matter in conformal models.
A proper $(\alpha,\beta)$-differential has to be invariant under
holomorphic changes of charts, thus induces a local rescaling by the
$\lambdan$'s,

\begin{eqnarray}
\phi_{(\alpha,\beta)}\ZBZn {(d\Zn)}^\alpha {(d\Zbn)}^\beta &=&
(\lambda^{\Zn}_z\zbz)^\alpha (\blambda^{\Zbn}_\bz\zbz)^\beta
\phi_{(\alpha,\beta)}\ZBZn\nn\\
&&\hskip -2.cm
\times\ ( dz+\mu(n,\zbz)d\bz)^\alpha (d\bz+\mub(n,\zbz)dz)^\beta 
\lbl{varphi}\\
&& \hskip -2.cm \equiv\
\varphi_{(\alpha,\beta)}\zbz  ( dz+\mu(n,\zbz)d\bz)^\alpha
( d\bz+\mub(n,\zbz)dz)^\beta \nn
\end{eqnarray}
 
The field will be said scalar if $(\alpha,\beta)=(0,0)$, namely,
\begin{eqnarray}
\phi\ZBZn
\equiv \varphi\zbz,\qquad\forall n 
\lbl{varphi1}
\end{eqnarray}
and will transform with only under point displacement
 \begin{eqnarray}
 \sS\phi(\Zn,\Zbn)= 
 \biggl(\Upsilon^{(n)}\prt_{\Zn}+\Upsilon^{\Zbn}\bprt_{\Zbn}\biggr)
 \phi(\Zn,\Zbn) 
\lbl{S-SCAL}
 \end{eqnarray}
 Going now  to background $\zbz$ we have 
\begin{eqnarray} 
\sS\varphi\zbz= 
 \biggr(\kappa_n^z\zbz\prt
  +\bar{\kappa_n}^\bz\zbz\bprt\biggl)\varphi\zbz
  \lbl{svarphi} 
\end{eqnarray}  
   
Now each $\ZBZn$ space has a different complex structure, so using the 
background representation each field   living in this space carries
 into its transformation the imprinting of this space.
 \begin{eqnarray}
\sS \phi_{(\alpha,\beta)}\ZBZn
&=& \biggl(\Upsilon^{(n)}\prt_{\Zn}+\Upsilon^{\Zbn}\prt_{\Zbn}\biggr)
\phi_{(\alpha,\beta)}\ZBZn \nn\\
&=& \biggr(\kappa_n^z\prt + \bar{\kappa_n}^\bz\bprt\biggl) 
\varphi_{(\alpha,\beta)}\zbz
\lbl{sphi} 
\end{eqnarray}
The same can be done with respect to the background system of
coordinates 

\begin{eqnarray}
\sS\varphi_{(\alpha,\beta)}\zbz &=&
(\kappa_n^z\prt+\bar{\kappa}_n^\bz\bprt)
\varphi_{(\alpha,\beta)}\zbz  \nn\\
&&+\ \alpha(\prt\kappa_n^z+\mu(n,\zbz)\prt\bar{\kappa}^\bz_n)
\varphi_{(\alpha,\beta)}\zbz 
\lbl{svarphi1} \\
&&+\ \beta(\bprt\bar{\kappa_n}^\bz+\mub(n,\zbz)
\bprt\kappa_n^z)\varphi_{(\alpha,\beta)}{(z^\alpha\bz^\beta)}\zbz\nn
\end{eqnarray}

In conclusion the above ghost decompositions Eqs\rf{kn} \rf{cA}
clarify our strategy towards a treatment of $w$ algebras in a 
Lagrangian Quantum Field Theory framework ; 
since from the former it is quite straightforward to derive in
combination with  the canonical 
construction of the diffeomorphism B.R.S. operator, the one induced by
the 
$w$ ordinary algebras. This will be very useful in the next Section.

The diffeomorphism variations of the "matter fields"
  $  \phi_{(\alpha,\beta)}\ZBZn$, fix, from our point of view ,
  their $w$ transformations, since it will be provided by 
  the decomposition of ghosts $\kappa_n^z\zbz$ in terms of the true
   $c^{(r,s)}\zbz$ symplectomorphism ghost fields.
   
 In particular, for the scalar field, the B.R.S. variation
Eq\rf{svarphi1}
 is rewritten as:
 
 \begin{eqnarray} 
\sS\varphi\zbz = \sum_{\scriptsize{ \begin{array}{c} r,s=0\\ r+s\geq
1\end{array}}}^n 
c^{(r,s)}\zbz\biggl(\tau^z_{n,(r,s)}\prt+\bar{\tau}^\bz_{n,(r,s)}\bprt\biggr)\varphi\zbz
\end{eqnarray} 
    
 We remark that this description is totally different from the the
 various approach to $w$ matter found in the literature
 e.g. \cite{Hull}. Moreover, according to this viewpoint, it gives
 completely trivial the problem.

\sect{Lagrangian Field Theory building}

 The approach we have given before to $w$ algebras, and in particular 
the 
 relevance of complex structures in their construction, suggests to
 investigate  
 the role played by these symmetries in a Lagrangian Field Theory.
 
 In the previous Sections we have emphasized the linking points 
 between $w$ algebras and two dimensional conformal
 transformations, so our discussion starts with an  
 example of conformal invariant models. 
  
 As it is  well known,  two dimensional conformal symmetry 
 means reparametrization invariance: in our $w$ scheme we have to 
 improve the symmetry of a wide class of reparametrization mappings,
 so a lot of care is required in order to respect the Lagrangian Field
Theory 
 prescriptions. 

We shall deal with a common conformal model built on a two dimensional
 space manifold $\cZ\zbz, \bar{\cZ}\zbz$.    

\bigskip

In order to construct a properly 
defined local Lagrangian theory we have to take care of:

1) a well definition of the Action integral, 

2) the locality on the constituent fields,

3) the symmetry constraints.  
 
We  consider the scalar field case:
\begin{eqnarray}
\leqa{ \Gamma_{scalar}=\int d\cZ \prt_{\cZ}
 \phi(\cZ,\bar{\cZ})\wedge d\bar{\cZ}\, \bprt_{\bar{\cZ}}
\phi(\cZ,\bar{\cZ})}\nn \\ 
&&\equiv 
\int d\Zn \prt_{\Zn} \phi\ZBZn \wedge d\Zbn \bprt_{\Zbn}
\phi\ZBZn               
\lbl{scalar1}
\end{eqnarray}

So we shall start from a model which is invariant under a
reparametrization
$\zbz \lra (\cZ\zbz,\bar{\cZ}\zbz)$, which is well defined but has quantum 
anomalies.

Our strategy for the realization of a $w$ symmetry in this model
will be to consider the $\cZ\zbz,\bar{\cZ}\zbz$ space  
as an $n$-{\em th} element of a $w$ space hierarchy as in Eq\rf{system}.

A positive answer for our purposes comes from Theorem \rf{guess3} but
more care 
has to be exercised.

For this reason the model is to be ``well defined" with respect 
all the possible backgrounds ; indeed 
the Lagrangian in  Eq \rf{scalar1} written  in terms of the  
$\zbz$ background  takes the form:

 \begin{eqnarray}
\Gamma_{scalar}&\equiv& \int dz\wedge d\bz\ \cL_{z,\bz}\zbz\nn\\
&=&\int dz\wedge d\bz\
\frac{\biggl[\prt-
\bar{\mu}(n,\zbz)\bprt\biggr]\phi\zbz 
\biggl[\bprt-\mu(n,\zbz)\prt\biggr]\phi\zbz}
{\biggl(1-\mu(n,\zbz)\bar{\mu}(n,\zbz)\biggr)}
\lbl{scalar2}
\end{eqnarray} 
 
Moreover the model is well defined in each  $ (Z^{(r)}, \bZ^{(r)})$
frame ($\forall r$) since: 
 \begin{eqnarray}
\Gamma_{scalar} &=&
\int d\Zn \prt_{\Zn} \phi\ZBZn\wedge d\Zbn \bprt_{\Zbn}
\phi\ZBZn  \nn \\
&=& \int dZ^{(r)}\wedge d\bZ^{(r)}
\biggl(1-\Xi_{\Zbr}^{\Zr}(n,\ZBZr)\bar{\Xi}_{\Zbr}^{\Zr}(n,\ZBZr)\biggr)^{-1}
\lbl{scalar3}\\
&& \hskip -2.5 cm \times 
\biggl[{\prt}_{Z^{(r)}} -
\bar{\Xi}_{\Zbr}^{\Zr}(n,\ZBZr){\bprt}_{\Zbr}\biggr]\phi\zbz 
\biggl[{\bprt}_{\Zbr}-\Xi_{\Zbr}^{\Zr}(n,\ZBZr){\prt}_{\Zr}\biggr]\phi\zbz\nn
\end{eqnarray}

This means that in this framework we can assume as symmetry 
transformations the changes of charts:
\begin{eqnarray}
\zbz \lra (\Zr\zbz, \Zbr\zbz)\quad r=1\cdots n\quad\forall n
\end{eqnarray}
just defined in Eq \rf{system}; 
and the dynamics of the  particle, which is free and scalar in  the
space 
$\ZBZn$, if described by means of the background 
of the underlying complex spaces $(\Zr\zbz,\Zbr\zbz\quad r=1\cdots n-1$
need the parametrization of the Beltrami multiplier $\mu_\bz^z(n\zbz)$
just found in the Eq \rf{mun}.

So at the light of  previous arguments and of Theorem
\rf{guess3} 

\begin{statement} 

 A two dimensional conformal 
model admits, at the classical limit, a $w$-symmetry of arbitrary order.

\end{statement}

Anyhow   the quantum extension requires some care.

Indeed  the $\lambda$'s,
are  non local functions of the $\mu_\bz^{z^s}\zbz$,so in a local
Lagrangian 
Quantum Field theory approach, they are not primitive, but,
they are essential for the geometrical meaning of our $w$ construction.

We have just seen in the preceding Lagrangian construction 
that, if we want to maintain the "well definition" of the Lagrangian 
with respect {\it{all}} the $\ZBZr$ frames, they are contained in the
Beltrami $\Xi_{\Zbr}^{\Zr}(n,\ZBZr)$
due to  Eqs \rf{xir} \rf{mun}. 

If we want to analyze how the underlying complex structure 
contributes to the dynamics the
price to pay is to put into the game all the 
$\lambda_z^{\Zr}$ fields, $ r=1,\cdots, n$ induced by the
decomposition of the $\ZBZn$-spaces Eq.\rf{Zdecomp}. These fields 
(even if local in the $\zbz$ background) are non local in the 
$\mu_\bz^{(r)}\zbz \quad r=1\cdots n$ fields due to the Beltrami
equations \rf{beltramin} in each $\ZBZr$, $ r=1,\cdots, n$ sectors.
So, according to this point of view, the model becomes 
{\it{intrinsically}} non local in the fields (unless $n=1$).

 We have now to choose the set of fields
which exhausts the dynamical configuration space :
so our coordinates will be the fields:  $\varphi$,
$c^{(r,s)}$, $\mu_\bz^{(r)}$ ,$ \lambda_z^{Z^r}$,
$r=1,\cdots, n$ and their derivatives.
This means that all the Classical B.R.S. diffeomorphism transformations
of these fields 
have to be written in terms of the $c^{(p,q)}$ ghosts using the
expansion of the various ghosts
$\cC^{(j)}$, $\cK^z_n$ and $\kappa^z_n$ as written previously.
 
So we define as ``naive " BRS functional operator the following $\delta_c$:
\begin{eqnarray}
\delta_{c} = \sum_{n=1}^{n_{max}}\int dz\wedge d\bz\biggl[
\biggl(\lambdan\zbz(\kappa_n^z
+\mu(n,\zbz)\bar{\kappa_n}^\bz\zbz\biggr)\frac{\delta}{\delta
\Zn\zbz} \nn \\
+ \sum_{\scriptsize{ \begin{array}{c} r,s=0\\ r+s\geq
1\end{array}}}^n \biggl(rc^{(r,s)}\zbz\prtz
c^{(p-r+1,q-s)}\zbz  
+ s c^{(r,s)}\zbz\prtbz c^{(p-r,q-s+1)}\zbz\biggr)
\frac{\delta}{\delta c^{(p,q)}\zbz}\nn\\
+\biggl(\bprt C^{(n)}\zbz - \sum_{r=1}^n\biggl
[ r\mu_{\bz}^{(r)}\zbz \prt C^{(n-r+1)}\zbz  
- r C^{(r)}\zbz \prt \mu_{\bz}^{(n-r+1)}\zbz\biggr] \biggr)
\frac{\delta}{\delta \mu_{\bz}^{(n)}\zbz}\nn\\ 
+ \biggl((\kappa_n^z\prt+\bar{\kappa_n}^\bz\bprt)\lambdan\zbz
+\lambdan\zbz(\prt\kappa_n^z+\mu(n,\zbz)\prt\bar{\kappa}^\bz) \biggr)
\frac{\delta}{\delta \lambdan\zbz}\nn\\
+\biggl[
\sum_{\scriptsize{ \begin{array}{c} r,s=0\\ r+s\geq
1\end{array}}}^n
c^{(r,s)}\zbz\biggl(\tau^z_{n,(r,s)}\prt+\bar{\tau}^\bz_{n,(r,s)}\bprt\biggr)\varphi\zbz\biggr]\frac{\delta}{\delta
\varphi\zbz } 
+c.c. \biggr]\nn\\ 
\lbl{deltac}
\end{eqnarray}
where both the ghosts $\kappa^z_n,\ \cC^{(j)}$ have been expressed in
terms of the $c^{(p,q)}$ ghosts according to Eqs.\rf{kappaA}\rf{Cc1}
respectively. 

This is the ordinary diffeomorphism BRS operator, and its 
 nilpotency is verified if the Beltrami conditions \rf{beltramin}
hold for all the $\lambda$'s \cite{BaLa}.
   
So the invariance of the Lagrangian $\Gamma_{Scalar}$ in Eq
\rf{scalar1} is written in a local form
\begin{eqnarray}
\delta_c\cL_{z,\bz}\zbz
=\prt(\kappa_n^z\zbz\cL_{z,\bz}\zbz)+\bprt(\bar{\kappa}_n^\bz\zbz\cL_{z,\bz}\zbz)
\end{eqnarray}

We now define a set of local operators of zero F-P charge by:
\begin{eqnarray}
\delta_{c}=\int dz\wedge d\bz \
\sum_{\scriptsize{ \begin{array}{c} p,q=0\\ p+q\geq
1\end{array}}}^{n_{max}} 
\biggl( c^{(p,q)}\zbz \cT_{{(p,q)}}\zbz  +
\cS c^{(p,q)}\zbz \frac{\delta}{\delta c^{(p,q)}\zbz} \biggr)
\lbl{tauc}
\end{eqnarray}
then thanks to both $\{\delta_c,\delta_c\}=0$, and Eq.\rf{0000a}, it
turns out that the $\cT_{{(p,q)}}\zbz$'s fulfill commutation rules of 
$w$-algebra type, see e.g. \cite{She92} and references therein : 
 \begin{eqnarray} 
\leqa{ \biggl[\cT_{(p,q)}\zbz, \cT_{(r,s)}\zbzp\biggr]
\ =}\nn\\
&&=\ p\, \prt_{z'}\delta^{(2)}(z'-z)\cT_{(p+r-1,q+s)}\zbz 
- r\, \prtz\delta^{(2)}(z-z') \cT_{(p+r-1,q+s)}\zbzp \nn\\[2mm]
&&+\ q\,\bprt_{z'} \delta^{(2)}(z'-z) \cT_{(p+r,q+s-1)}\zbz
- s\, \prtbz \delta^{(2)}(z-z') \cT_{(p+r,q+s-1)}\zbzp.
\lbl{02}
\end{eqnarray}
Note however that the obtained $w$-algebra with respect to
the $c^{(p,q)}$ ghosts is not chiral, but in the vacuum sector where the one
relative to the $C^{(j)}$ ghosts is chiral.
  
 \begin{statement}
 The ordinary diffeomorphism BRS transformations will induce,
 from Eqs\rf{kn} \rf{cA}  $w$-algebra symmetry transformations. 
 The BRS functional operator to be used for the Field Theory quantum 
 extension the diffeomorphism symmetry (in terms of the
 $\cK_n,\kappa_n$ ghosts),
 will give, if written in terms of the $c^{(r,s)}$ ghosts,
 a BRS differential for $w$-algebras.  
 \end{statement}

The ordinary procedure for Quantum extension suggests  to introduce
the anti-fields in the Lagrangian term: 

\begin{eqnarray}
\cL_{antifields} = \int dz\wedge d\bz \biggl(\sum_{r,s}
 ( \xi_{(r+1,s+1)}\zbz \sS 
 c^{(r,s)}\zbz)
 +\nu_{(s+1)}\zbz \sS\mu_\bz^{(s)}\zbz \nn \\
  + \chi_{z,\bz}\zbz \sS \varphi\zbz +\sum_r
({\rho(r,\zbz)}_{\bz,{\Zr}}\sS
 \lambda^{(\Zr)}_z\zbz) + c.c.\biggr)
 \end{eqnarray} 
So the complete Classical Action becomes  
\begin{eqnarray}
\Gamma^{Classical}=\Gamma_{scalar}+\Gamma_{antifields}
\end{eqnarray} 
 
and at the classical level we get:

\begin{eqnarray}                                                 
\delta \Gamma^{(Classical)}\equiv \int dz\wedge d\bz\biggl[
\frac{\delta \Gamma^{(Classical)}}{\delta \chi_{z,\bz}\zbz} 
\frac{\delta \Gamma^{(Classical)}}{\delta \phi\zbz}+
\frac{\delta \Gamma^{(Classical)}}{\delta \xi_{(r+1,s+1)}\zbz}
\frac{\delta \Gamma^{(Classical)}}{\delta c^{(r,s)}\zbz}\nn \\[2mm]
+\frac{\delta \Gamma^{(Classical)}}{\delta \nu_{(s+1)}\zbz}
\frac{\delta \Gamma^{(Classical)}}{\delta \mu_{\bz}^{(s)}\zbz}
+\frac{\delta \Gamma^{(Classical)}}{\delta {\rho(r,\zbz)}_{\bz,\Zr}}  
\frac{\delta \Gamma^{(Classical)}}{\delta \lambda_{z}^{\Zr}\zbz}
+ c.c. \biggr]=0 \nn \\
\lbl{brs}
\end{eqnarray}

By the way if we want to reproduce here one of the outstanding  feature
of conformal models
, that is the holomorphic properties of the object coupled in an
invariant way to Beltrami fields (i.e. the 
Energy Momentum tensor) the task is not so simple.

This fact is, in this context, particularly fruitful :
the presence of $n$ independent complex structures (and then $n$ independent
Beltrami fields) means that we can derive at least $n$ energy-momentum tensors 
and their related holomorphic properties.

The problem is that the Beltrami multipliers are non local objects,
so the Energy-Momentum tensor cannot be defined in terms of 
functional derivatives except for the case $n=1$.

 We can provide a solution by the following shortcut
 :  introduce 
 the following lower triangular $n_{max}\times n_{max}$ matrix $\cA$
 with entries 
 $(r-1,0)$-differentials valued bilocal kernels - but highly non local
 in the $\mu_\bz^{(j)}$'s,  
\begin{eqnarray} 
\cA(n,r;\zbz,\zbzp)_{(r-1)}
\equiv\frac{\delta\mu_\bz^z(n,\zbz)}{\delta \mu_\bz^{(r)}\zbzp},\qquad
n=1\cdots n_{max},\ 0\leq r\leq n,
\end{eqnarray}
such that (compare with the expansion \rf{mun}),
\be
\mu_\bz^z(n,\zbz) = \int dz'\wedge d\bz' \sum_{r=1}^n 
\cA(n,r;\zbz,\zbzp)_{(r-1)} \mu_\bz^{(r)}\zbzp.
\ee
We shall suppose that $\cA$ has an inverse $\cB$ with entries
 $(2-r,1)$-differentials valued bilocal kernels, such that everywhere,
\begin{eqnarray}
\int dz"\wedge d\bz" \sum_{r=1}^{n_{max}} \cB(n,r;\zbz\zbzpp)_{(2-r,1)}
\cA(r,n';\zbzpp,\zbzp)_{(r-1,0)} = \nn\\
= \delta_{n,n'}\delta^{(2)}(z-z').
\end{eqnarray}

If we define: 
\begin{eqnarray}
\cP_{z}^{\bz}(n,\zbz)\equiv \int dz'\wedge d\bz'
\sum_r \cB(n,r;\zbz\zbzp)_{(2-r,1)}\frac{\delta}{\delta \mu^r\zbzp}
\lbl{inverse}
\end{eqnarray} 
one thus has 
\begin{eqnarray}
\cP_{z}^{\bz}(n,\zbz) \mu(n'\zbzp)
=\delta_{n,n'}\delta^{(2)}(z-z').
\lbl{derivative}
\end{eqnarray} 

The $\cP$'s  play the role of the 
"functional derivative operators" with respect the 
Beltrami parameters; they will be (as well as these last) 
non local (in $\mu^{(r)}\zbz$ ) functional operators and have 
a fundamental role in our context.
If $\mu(n,\zbz)$is coupled at the tree approximation
 with a local  field $\Theta^{(Classical)}_{(zz)}(n,\zbz)$ in an
invariant way,
for each n $n=1\cdots n_{max}$ we have
\begin{eqnarray} 
 \Theta^{(Classical)}_{(zz)}(n,\zbz) =\cP_{z}^{\bz}(n,\zbz)
\Gamma^{(Classical)}
 \end{eqnarray} 
so that the latter will transform at the classical level as:
\begin{eqnarray}
\sS \Theta^{(Classical)}_{(zz)}(n,\zbz)= \cK^n \zbz\prt
\Theta^{(Classical)}_{zz}\zbz + 2 \Theta^{(Classical)}_{zz}\zbz\prt
\cK^n\zbz
\end{eqnarray}
By the anticommutator ${\dps \bprt = \{ \cS, \frac{\prt}{\prt \kappa_n}
\}}$
one gets :
  \begin{eqnarray}
\biggl(\bprt-\mu(n,\zbz)\prt-2\prt
\mu(n,\zbz)
\biggr) \Theta^{(Classical)}_{(zz)}(n,\zbz)=0
\end{eqnarray}
Defining the $(2,0)$-differential
\begin{eqnarray}
\cJ_{Z^{(n)}Z^{(n)}}^{(Classical)}(n,\ZBZn)\equiv 
\biggl[
\frac{1}{{\lambdan}^2}\Theta^{(Classical)}_{(zz)}(n,\zbz)\biggr] 
\end{eqnarray} 

the previous equation leads to:
\begin{eqnarray}
\frac{\prt}{\prt{\bar{Z}}^n}
\cJ_{Z^{(n)}Z^{(n)}}^{(Classical)}(n,\ZBZn) 
=0 
\end{eqnarray}

 In particular we remark that in the $\zbz$ background:
\begin{eqnarray}
d  \cJ_{Z^{(n)}Z^{(n)}}^{(Classical)}(Z^{(n)}\zbz) =\biggl[\frac{\prt
\cJ_{Z^{(n)}Z^{(n)}}^{(Classical)}(Z^{(n)}) }{ \prt Z^{(n)}}\biggr]\zbz
\lambdan\zbz\biggl[ dz +\mu(n,\zbz) d\bz\biggr]\nn\\
\lbl{tensorcomplexclass} 
\end{eqnarray}

This means that:
\begin{statement}

The conserved current $\cJ_{Z^{(n)}Z^{(n)}}^{(Classical)}$ 
is a non local function of $\mu(n,\zbz)$. 
It will imply that this object is a nonlocal function of
 $\mu_{\bz}^{(j)}$ for $j\leq n$.
\end{statement} 
 
 Moreover we can rewrite the B.R.S transformation of the current as:
 \begin{eqnarray}
 \sS
\cJ_{Z^{(n)}Z^{(n)}}^{(Classical)}(Z^{(n)})=\Upsilon^{n}\prt_{Z^{(n)}}
  \cJ_{Z^{(n)}Z^{(n)}}^{(Classical)}(Z^{(n)})
  \end{eqnarray}
  
 so that we can define a set of invariant charges:
  \begin{eqnarray}
  \cQ^{(Classical)}_n=\int
  \cJ_{Z^{(n)}Z^{(n)}}^{(Classical)}(Z^{(n)})d Z^{(n)}   
  \end{eqnarray}
  
  which are functional depending on the local parameters
  $\mu_\bz^{(j)}\zbz$ for $j\leq n$,  
   
  \begin{eqnarray}
  \sS\cQ^{(Classical)}_n =0,\qquad    \forall n= 1\cdots n_{max}       
  \end{eqnarray}
  
and {\em a fortiori}:  
\begin{eqnarray}
\cT_{(p,q)}\cQ^{(Classical)}_n =0, \qquad
\forall p,q\leq n =1\cdots n_{max},         
\end{eqnarray}
that is the  charges $\cQ_n$ are invariant under both diffeomorphism
and $w$ action.

Even if we have stressed the "non local" nature of our 
theory, we can ask whether  some noteworthy property is hidden in 
the pure local sector of the model.

The local counterpart of the Energy-momentum tensor (which 
is  invariantly coupled to the Beltrami fields) are the
quantities 
which are coupled in an invariant way to the $ \mu_\bz^{(j)}$'s.

We have already pointed out that in this context all the geometrical
architecture 
of our building  cannot be appreciated;anyhow a relic of $w$ 
algebras still appears : we now show, 
that their OPE's will generate a $w$ expansion,as it has already be
shown in the Literature \cite{4}. 
Indeed introducing, at the Classical level,
the Ward identity for the appropriate partition function 
$\cZ^{(Classical)}(\mu)$ of the vacuum sector is:

\begin{eqnarray}
\bprt\frac{\delta\cZ^{Classical}(\mu)}{\delta
\mu_{\bz}^{(s)}\zbz}
-\sum_{j= 0}^{n_{max}-s}\biggl((s+j+1)\prt\mu_{\bz}^{(j+1)} 
+(j+1)\mu_{\bz}^{(j+1)}\prt\biggr)
\frac{\delta\cZ^{Classical}(\mu)}{\delta
\mu_{\bz}^{(j+s)}\zbz}
=0 \nn\\
\lbl{wardmus}
\end{eqnarray}
from which we derive by multiplying by $\pi\frac{1}{(z-z')}$ and
integration,
\begin{eqnarray} 
\pi^2\frac{\delta\cZ^{Classical}(\mu)}{\delta
\mu_{\bz}^{(s)}\zbzp}+
\sum_{j= 0}^{n_{max}-s}\int dz\wedge d\bz\ \mu_{\bz}^{(j+1)}\zbz  
\biggl(\frac{s+j+1}{{(z-z')}^2} 
+\frac{s}{(z-z')}\prt\biggr)\pi
\frac{\delta\cZ^{Classical}(\mu)}{\delta
\mu_{\bz}^{z^{(j+s)}}\zbz}=0\nn \\
\end{eqnarray} 
Setting all the $\mu^{(j)}_\bz$'s to zero and by quantum action
principle one thus gets 
 
 \begin{eqnarray} 
\left.\left.\pi^2\frac{\delta\cZ^{Classical}(\mu)}{\delta
\mu_{\bz}^{(s)}\zbzp\delta\mu_{\bz}^{(r)}\zbz }\right|_{\mu=0} +
\sum_{j= 0}^{n_{max}-s} \delta^{j+1}_r  \biggl(\frac{s+r}{{(z-z')}^2} 
+\frac{s}{(z-z')}\prt\biggr)\pi
\frac{\delta\cZ^{Classical}(\mu)}{\delta
\mu_{\bz}^{(r+s-1)}\zbz}\right|_{\mu=0} =0\nn\\
\end{eqnarray} 
which gives the OPE for the tensors coupled to these objects.

This is valid only at the classical level: the local theory 
display at the quantum level anomalies, while the "non local" 
approach admits, as we shall see in the next Section, a rather 
painless cancellation mechanism.

 \subsect{Quantum extension and Anomalies}

The difficulties  avoided using  
a local $w$ algebra  using $\cC^{(n)}\zbz$ or $c^{(r,s)}\zbz$ ghosts
will create 
other problems for the Quantum extension of the model.

Due to quantum  perturbative corrections the Action functional may
violate the Ward identities, and according to the usual Lagrangian
framework, one introduces the corresponding linearized BRS operator,
 
\begin{eqnarray}
\delta \Gamma=\Delta
\end{eqnarray} 
\begin{eqnarray}                                                 
\delta \equiv \int dz\wedge d\bz\biggl[
\frac{\delta \Gamma^{(Classical)}}{\delta \chi_{z,\bz}\zbz} 
\frac{\delta }{\delta \phi\zbz}+
\sum_{r,s}\biggl(\frac{\delta \Gamma^{(Classical)}}{\delta
\xi_{(r+1,s+1)}\zbz}
\frac{\delta }{\delta c^{(r,s)}\zbz}\biggr)\nn \\
+\sum_{s}\biggl(\frac{\delta \Gamma^{(Classical)}}{\delta
\nu_{(s+1)}\zbz}
\frac{\delta }{\delta \mu_{\bz}^{(s)}\zbz}\biggr)
+\sum_{r}\biggl(\frac{\delta \Gamma^{(Classical)}}{\delta
\rho_{z,\Zr} (r,\zbz)}   
\frac{\delta }{\delta \lambda_{z}^{\Zr}\zbz}\biggr)\nn \\
+\frac{\delta \Gamma^{(Classical)}}{\delta \phi\zbz}
\frac{\delta }{\delta \chi_{z,\bz}\zbz}
+ \sum_{r,s}\biggl(
\frac{\delta \Gamma^{(Classical)}}{\delta c^{(r,s)}\zbz}
\frac{\delta }{\delta \xi_{(r+1,s+1)}\zbz}\biggr)\nn \\
+ \sum_s\biggl(
\frac{\delta \Gamma^{(Classical)}}{\delta \mu_{\bz}^{(s)}\zbz}
\frac{\delta }{\delta \nu_{(s+1)}\zbz}\biggr) 
+ \sum_{r}\biggl( 
\frac{\delta \Gamma^{(Classical)}}{\delta \lambda_{z}^{\Zr}\zbz}
 \frac{\delta }{\delta \rho_{z,\Zr}(r,\zbz)} + c.c.\biggr) 
\biggr]
\lbl{brslin}
\end{eqnarray}    
so only by counter-terms inclusion the symmetry will be restored at each
order 
of the perturbative expansion. As is well-known, this requires a
cohomological approach and if the cohomology is empty, the   
symmetry is restored at the quantum level.

This  calculation is performed in the Appendix,
where we show that the cohomology sector in the space of the local
functions 
is isomorphic to the tensor product of the cohomologies 
of the ordinary disjoint smooth changes of coordinates
$\zbz\lra (\Zr\zbz,\Zbr\zbz),\ \forall r=1\cdots n$.
 
This result shifts the problem to the quantum extension of 
a theory whose symmetry is provided by $n$ disjoint ordinary 
 changes of coordinates, for which many known results are at our disposal 
\cite{BaLa}\cite{BaLa1}

In particular if we add to our field content {\it {all}} the 
$\ln\lambda^{Z(r)}\zbz\quad r=1\cdots n_{max}$ the cohomology 
becomes empty.

The implicit``non locality on the fields" of our model softens 
the possible disappointment coming from the introduction of logarithms.

In the usual Quantum Field Theory the local anomaly is a cocycle which
has a coboundary 
term which is $\log\lambda$ dependent and derives from the usual
transgression formulas 
coming from the Gel'fand-Fuchs cocycle, which becomes coboundary if we put
$\ln\lambda$'s into 
the game: in our case we get: 
 
 \begin{eqnarray}
 \Delta^\natural\zbz &=&\sum_{r=1}^n c_r\cK_r\zbz\prt\cK_r\zbz\prt^2
 \cK_r\zbz\nn\\ 
  &=&\sum_{r=1}^n
c_r\,\delta\biggl(\cK_r\zbz\prt\cK_r\zbz\prt\ln\lambda_{z}^{(Z^r)}\zbz\biggr) 
  \lbl{anomaly} 
  \end{eqnarray} 
  modulo  coboundary terms and total derivatives; the anomaly in the
 space of local functionals is recovered by using the techniques of Ref
 \cite{BaLa,Lazn}.  
\begin{eqnarray}
\Delta\zbz=\sum_{r=1}^{n_{max}} c_r\, \cK_r^z\zbz \prt^3
          \mu_{\bz}^z(r,\zbz)\nn\\ 
          \end{eqnarray}

Anyhow the B.R.S diffeomorphism operator  is deeply related to the one
of $w$ 
symmetry when we render explicit the $\cK_n\zbz$ ghosts in terms of
$c^{(r,s)}\zbz$ ones;
this allows to calculate the $w$ local anomalies: 
 \begin{eqnarray}
\cT_{(r,s)}\zbz\Gamma= \sum_{n=\mbox{\scriptsize max}(r,s)}^{n_{max}}
 c_n\, \eta_{(r,s)}(n,\zbz)\prt^3 \mu(n,\zbz).
\end{eqnarray}

 From the usual construction \cite{BaLa}
 
 \begin{eqnarray}
 \cK_r\zbz\prt\cK_r\zbz\prt^2\cK_r\zbz=
 \delta( \cK_r\zbz\prt\cK_r\zbz\prt\ln \lambda_{z}^{\Zr}\zbz) 
 \end{eqnarray}

 so the non local Action compensating terms will be:  
\begin{eqnarray}
\Gamma^{(Polyakov)}=
\sum  c_r \int dz\wedge d\bz (\mu_{\bz}^z(r,\zbz)
\prt^2 \ln\lambda^{\Zr}\zbz 
\nn \\
\end{eqnarray}
and the corresponding e-m tensor
\begin{eqnarray} 
 \Theta^{(Polyakov)}_{(zz)}(n,\zbz)
 =\cP_{z}^{\bz}(n,\zbz)\Gamma^{(Polyakov)}= -2 c_n\, S_{zz}(\Zn\zbz) 
 \lbl{thetapolyakov} 
  \end{eqnarray} 
 
where, as usual:
\begin{eqnarray}
  S_{zz}(Z^{(n)}\zbz)\equiv 
\prt^2
 \ln \lambdan\zbz -\frac{1}{2}\biggl(\prt\ln \lambdan \zbz\biggr)^2
\end{eqnarray}

 So we can define:
  
 \begin{eqnarray} 
 \Theta_{(zz)}(n,\zbz) =\cP_{z}^{\bz}(n,\zbz)\biggl[
 \Gamma-\Gamma^{(Polyakov)} \biggr] 
\end{eqnarray}

 So the anomaly compensation mechanism allows 
 to construct at the Quantum level an energy momentum tensor 
 which verifies the same symmetry properties as the classical energy
 momentum tensor.  
    
 At this stage it is trivial to define a current
 \begin{eqnarray}
\cJ_{Z^{(n)}Z^{(n)}}(n,\ZBZn) \equiv
 \frac{1}{{\lambdan}^2}\biggl[\cP_{z}^{\bz}(n,\zbz)
  \Gamma
+2\, c_n\, S_{zz}(Z^{(n)}\zbz) \biggr]
\end{eqnarray}
which is the Quantum extension of the classical $(2,0)$-covariant tensor
$\cJ_{Z^{(n)}Z^{(n)}}^{(Classical)}(n,\ZBZn)$. Note that due to  the
presence of the Schwarzian derivative the former is no longer a tensor. 

Moreover we can also define, for each $n\leq n_{max}$:

\begin{eqnarray}
\cQ_n=\int  \cJ_{Z^{(n)}Z^{(n)}}(Z^{(n)})d Z^{(n)}  
  \end{eqnarray}

which will be invariant even in the Quantum level.

 \sect{Conclusions} 
 
The many  aspects of two dimensional reparametrization invariance 
provide a further geometrical description of $w$-algebras. We have
addressed the question 
of introducing local $(-n,1)$-conformal fields generalizing
the usual Beltrami differential appearing in $w$-gravity. 
It was shown that the way out is based on the infinitesimal action of
symplectomorphisms on coordinate transformations
dictated by very special canonical transformations.

Also, it is both interesting and intriguing to note how intermingled
the symplectic and conformal geometries are relevant for all the
present treatment. The combination of Beltrami parametrization of
complex structures, canonical transformations and symplectomorphisms
yields to a BRS formulation of $w$-algebras.

However, although the locality requirements are fundamental for the
physical  
contents within a Lagrangian field Theory, we have overcome them in
order to take ever present the geometrical aspect of the problem. But we aim
to treat the former in order to understand better the role of the
Quantum $w$ local 
anomalies \cite{Hull,Sorella,Grimm,Ader1} in relation to the point of
view expressed in the present paper. 

\indent

\noindent
{\bf Acknowledgements}. We are grateful to Prof. A. Blasi for comments
and discussions.  

\sect{Appendix}

\indent

The purpose of this Appendix is to show that the cohomology space 
of our BRS operator  $\delta$ in the space of local functions,
 is isomorphic to the one of the $n$ independent reparametrizations
$\zbz\lra\ZBZn$.
   
We have shown in \cite {BaLa} that the cohomology space in the functional
of the BRS operator $\delta$ will coincide with the local function
cohomology of the nilpotent BRS operator $\delta
-c^{(1,0)}\prt-c^{(0,1)}\bprt$.

This cohomology space will be computed by using the spectral sequences
method. Let us filter with:
\begin{eqnarray}
\nu=\sum_{p,q,m,n} {\biggl(p+q\biggr)} \prt^m\bprt^n c^{(p,q)}\zbz
\frac{\prt}{\prt   \prt^m\bprt^n c^{(p,q)}\zbz }.
\lbl{counter}
\end{eqnarray} 
At the zero eigenvalue the following operator is obtained,

\begin{eqnarray}                                                 
\delta_0 \equiv \int dz\wedge d\bz\biggl[
 \frac{\delta\Gamma^{(Classical)} }{\delta \phi\zbz}\frac{\delta
}{\delta \chi_{z,\bz}\zbz}
+\sum_s \biggl(\frac{\delta \Gamma^{(Classical)}}{\delta
\mu_{\bz}^{(s)}\zbz}
\frac{\delta }{\delta \nu_{(s+1)}\zbz}\biggr) \nn \\
\left. +\sum_{r,s}\biggl(\frac{\delta \Gamma^{(Antifields)}}{\delta
c^{(r,s)}\zbz} \right|_{c =0} 
\frac{\delta }{\delta \zeta_{(r+1,s+1)}\zbz}\biggr) 
+
\sum_r \biggl(\frac{\delta \Gamma^{(Classical)}}{\delta
 \lambda_{z}^{\Zr}\zbz}\frac{\delta }{\delta \rho_{z,\Zr}(r,\zbz)}
\biggr)   
\biggr]
\lbl{brs0}
\end{eqnarray} 
where ${\dps \frac{\delta \Gamma^{(Antifields)}}{\delta c^{(p,q)}\zbz}
\arrowvert_{c =0}}$ is the $c$ independent part  of the BRS variation
of the anti-fields $\zeta$ induced by the linearization of $\delta$.

This operator is clearly nilpotent due to the $\Phi$-$\Pi$ neutrality of
the $\Gamma^{Classical}$ terms. Its cohomology space can be
calculated using again the spectral sequences method.
Its adjoint can be defined upon using the Dixon procedure \cite{Dixon}
and the Laplacian kernel is isomorphic to the cohomology space. 
 The upshot of this calculation 
does not modify the final result; anyhow for the sake of completeness we can
calculate this space by first filtrating this operator 
with the field operator counter, 
and by calculating the kernel of the Laplacian. It is easy to convince
one self that the cohomology space 
will be independent on the anti-fields $ \rho_{z,\Zr}(r,\zbz)$,$
\nu_{(s+1)}\zbz$, $\chi_{z,\bz}\zbz$, $\zeta_{r+1,s+1}\zbz$ and
complicated combinations in the
matter fields and $\lambda$'s and $\mu$'s. 

The fundamental step takes place in the analysis of the action on
$\delta$
of the filtering operator \rf{counter} at the eigenvalue equal to one.
In this case we have to calculate the kernel of this operator (and its
adjoint) on the space previously calculated with $\delta_0$ .
It is easy to derive that this operator is nothing else but the sum 
of the operators $\delta -c^{(1,0)}\prt-c^{(0,1)}\bprt$, 
(where $\delta$ is the ordinary 
diffeomorphism operator of the $\Phi$-$\Pi$ neutral fields
 containing the ghosts $c^{(1,0)}$ and $c^{(0,1)}$)
 plus the total variation of $c^{(p,q)}$.

This operator is still nilpotent so we can filter it again.
We shall choose as filtering operator the one which counts the
$c^{(1,0)}$ and $c^{(0,1)}$ ghost fields, namely, 
\begin{eqnarray}
\nu^{\prime}=\sum_{p,q,m,n}  \prt^n\bprt^m c^{(1,0)}\zbz
 \frac{\prt}{\prt   \prt^n\bprt^m c^{(1,0)}\zbz }+
  \prt^n\bprt^m c^{(0,1)}\zbz
 \frac{\prt}{\prt   \prt^n\bprt^m c^{(0,1)}\zbz }
 \lbl{counter1}
\end{eqnarray} 

At zero eigenvalue we find:
\begin{eqnarray}
\delta^{\prime}_0
=\sum_{\scriptsize \begin{array}{c} j,l,m,n,r,s,p,q\\
p+q>r+s>1 \end{array}}\sm{n!m!}{l!j!(n-l)!(m-j)!}\biggl[\prt^l\bprt^j 
c^{(r,s)}\zbz 
\biggl(\,r\prt^{(n-l+1)}\bprt^{(m-j)}c^{(p-r+1,q-s)}\zbz\nn\\
\hskip -2.cm +\ s\,\prt^{(n-l)}\bprt^{(m-j+1)}c^{(p-r,q-s+1)}\zbz\biggr)\biggr]
\frac{\prt}{\prt(\prt^n\bprt^m c^{(p,q)}\zbz)}
\lbl{scccc} 
\end{eqnarray}  
After defining its adjoint according to the Dixon procedure it is easy
to find that the cohomology does not depend on  the ghost fields 
$ c^{(p,q)}$ and their derivatives, with the condition
$\biggl(p+q\biggr)>1$.

At the end we are left with the BRS operator induced by the
following transformation rules, for any $n$: 
 
\begin{eqnarray}
\sS\mu^{(n)}_\bz\zbz = c^{(1,0)}\zbz\prt\mu^{(n)}_\bz\zbz 
+\bprt\biggl(\mu^{(n)}_\bz\zbz c^{(0,1)}\zbz\biggr)+\bprt c^{(1,0)}\zbz
\,\delta_{n,1}\nn\\
-n \mu^{(n)}_\bz\zbz \prt c^{(1,0)}\zbz - \biggl(
\sum_rr\,\mu_\bz^{(r)}\zbz  
\mu^{(n-r+1)}_\bz\zbz\biggr) \prt c^{(0,1)}\zbz\nn\\
\lbl{smuc} 
\end{eqnarray}
 
\begin{eqnarray}
\sS\lambdan\zbz &=& \biggl( c^{(1,0)}\zbz \prt+ c^{(0,1)}\zbz \bprt\biggr)
\lambdan\zbz\nn\\
&&+\ \lambdan\zbz \biggl(\prt c^{(1,0)}\zbz  
+\mu(n,\zbz)\prt c^{(0,1)}\zbz\biggr) 
\lbl{slambdanc}
\end{eqnarray}
where $\mu(n,\zbz)$ must be written according to the expansion Eq.\rf{mun},
  we thus get the transformations for the Beltrami differentials at
  any level $n$:
 \begin{eqnarray}
\sS\mu(n,\zbz)= \biggl( c^{(1,0)}\zbz\prt+c^{(0,1)}\zbz\bprt
  \biggr)\mu(n,\zbz)\nn\\ 
+\ \bprt c^{(1,0)}\zbz+\mu(n,\zbz)\bprt c^{(0,1)}\zbz\nn\\
-\ \mu(n,\zbz) \biggl(\prt c^{(1,0)}\zbz+\mu(n,\zbz)\prt c^{(0,1)}\zbz
\biggr)
\lbl{smuc1} 
\end{eqnarray}
while for the scalar matter field,
 \begin{eqnarray}
\sS\varphi\zbz= \biggl(c^{(1,0)}\zbz\prt+c^{(0,1)}\zbz\bprt \biggr)
\varphi\zbz
\lbl{svarphi2} 
\end{eqnarray} 
 and the ghost field $c^{(1,0)}$ (and the c.c. for $c^{(0,1)}$),
\begin{eqnarray}
\sS c^{(1,0)}\zbz= \biggl(c^{(1,0)}\zbz\prtz
+  c^{(0,1)}\zbz\prtbz\biggr) c^{(1,0)}\zbz.
\lbl{0000c}
\end{eqnarray}
All of these are ordinary diffeomorphism transformations.

So the $w$ algebra reduces to a tensor product of
$n$ independent diffeomorphisms of level equal to one    
 $\zbz \lra \ZBZn,\ \forall n=1,\cdots, n_{max}$.


\begin{thebibliography}{10}
 
  
 \bibitem{Zam} A.B. Zamolodchikov, Teor. Mat. Fiz. 65, 1205 (1985)
 
 \bibitem{DrinSok} V.Drinfeld and V.Sokolov, J. Sov.Math, 30 ,1975 (1984)
 
 \bibitem{GelDik} I.M. Gel'fand, L.A. Dickey, Funct. Anal. Appl. 10,
 4,(1976); 11, 2,(1976) 
  
 \bibitem{4} A. Gerasimov, A. Levin and A. Marshakov, Nucl. Phys. B360,
 537(1991);

A. Bilal, V.V. Fock and I.I. Kogan, Nucl. Phys. B359,635 (1991);

 A. Das, W-J. Huang and S. Roy, Int. J. Mod. Phys. A7,3447(1992);

 J. de Boer and J. Goeree, Nucl. Phys. B381,329 (1992); Phys.Lett. B274,289
(1992)
 
K.Yoshida, Int. J. Mod. Phys. A7,4353(1992)
  
\bibitem{7} C. Itzykson, "W-geometry" , Cargese Lectures,  in
``Random Surfaces and  
 Quantum Gravity", ed. O.Alvarez et al, Plenum Press, New York, 1991
 
 \bibitem{7a} Y.Matsuo, Comm. Math. Phys. 152,317 (1993) and Phys. Lett
 B274,309 (1992)

 \bibitem{govin} S. Govindarajan, "Higher dimensions uniformisation
 of $w$ geometry", hep-th 9412078
  
\bibitem{Hit}N.J. Hitchin, Topology 31,451 (1992) 

\bibitem{Wi1}
E. Witten, "Surprises with topological field theories"
in "String 90", Proceedings, Superstring workshop,College Station USA, 
March 12-17, 1990,by R. Arnowitt and al. eds 
World Scientific Publishing, Singapore 1991.

\bibitem{Hull}
C.M.Hull, "Geometry and W-gravity", Talk at "Pathways to Fundamental
Interactions",
16th John Hopkins Workshop on Current Problems in Particle
Theory. Goteborg 1992, hep-th/9301074.

C.M. Hull, ``Classical and quantum $W$-gravity", hep-th/9201057.

C.M. Hull, ``$W$-Geometry", Comm. Math. Phys. {\bf 156}, 1973, 245-275, 
hep-th/9211113.

\bibitem{BaLa0} G. Bandelloni, S.Lazzarini, "$w$-algebras from canonical
tranformations", submitted for publication.
 \bibitem{BPZ} 
A.A. Belavin, A.M. Polyakov, A.B. Zamolodchikov, "Infinite 
conformal symmetry in two dimensional quantum field theory", Nucl. Phys.
B241 (1984) 333
\bibitem{BK}
A.A. Belavin, V.G. Knizhnik, Phys. Lett. B168, 201(1986), 
Sov. Phys. JEPT 64,214 (1986)
 
\bibitem{deBoerGoeree}
J.De Boer, J. Goeree, "Covariant W-gravity and its moduli space from
gauge theory",
Nucl. Phys. B401, 369 (1993), hep-th/9206098

\bibitem{becchi}
C.M. Becchi, Nucl.Phys.B 304 (1988) 513 
\bibitem{Zucchini}
R.Zucchini, "Light cone $W_n$ geometry and its symmetries 
and projective field theory" Class.Quant.Grav.10,253-278 (1993)
\bibitem{24}
G.Sotkov, M.Stashnikov and C.J.Zhu, Nucl.Phys B356, 245,(1991)
\bibitem{25}
G.Sotkov, M.Stashnikov,  Nucl.Phys B356, 439,(1991)
\bibitem{26}
J.L. Gervais, Y. Matsuo, Phys Lett B274, 309,(1992) 
\bibitem{Laz1}
S. Lazzarini, "Flat complex vector bundles, the Beltrami differentials, 
and W algebras", Lett. Math. Phys 41, 207-225, (1997)

\bibitem{Laz2}
S.Lazzarini,
 "Some remarks on the geometry of $W$-algebras", in
``$W$-algebras: Extended Conformal Symetries'', 
Marseille-Luminy, 3-7 July 1995, R. Grimm et V. Ovsienko Eds,
Preprint CPT-95/P.3268.
 
\bibitem{Gieres} F.Gieres, "Conformally covariant operators on Riemann 
Surfaces (with applications to conformal and integrable models)",
Int. J.M od. Phys. A8,1-58,1993

\bibitem{Sorella}
M. Carvalho, L.C. Queiroz Vilar, S. Sorella, "Algebraic Characterization of 
anomalies in chiral W(3) gravity", Int. J. Mod. Phys. A10, 3877-3900 (1995)


\bibitem{Grimm}
D. Garajeu, R. Grimm, S. Lazzarini, "W gauge structure and their
anomalies: an algebraic approach", J. Math. Phys 36, 7043-7072 (1995)

\bibitem{She92} X. Shen, ``W infinity and string theory",
Int. J. Mod. Phys. {\bf A7} (1992) 6953-6993.

\bibitem{Ader1}
A.A bud,J.-P. Ader, L. Cappiello, "Consistent anomalies of the induced W
gravities", Phys. Lett. B396 108-116 (1996)

\bibitem{Ader2} 
J.-P. Ader, F. Biet, Y. Noirot,
"A geometrical approach to super W-induced gravities in two dimensions",
Nucl. Phys. B466, 285-314 (1996)

\bibitem{BaLa}
G. Bandelloni, S. Lazzarini, "Diffeomorphism 
cohomology in Beltrami parametrization", J. Math. Phys.
 34 5413-5440 (1993)
\bibitem{BaLa1}
G. Bandelloni, S. Lazzarini, "Diffeomorphism 
cohomology in Beltrami parametrization: the 1 forms",  J. Math. Phys.
36 1-29 (1995)
\bibitem{Lazn}
S. Lazzarini, Phys. Lett. B436, 73 (1998)
\bibitem{Dixon}
J. Dixon: "Cohomology and Renormalization of Gauge Theories,I,II,III"
Unpublished reports.  
\end{thebibliography}
\end{document}